\documentclass[11pt]{article}
\usepackage{amsmath,amssymb,amscd}
\usepackage{amsthm,bbm}
\usepackage{mathtools}
\usepackage{float}
\usepackage[caption = false]{subfig}
\usepackage[final]{graphicx}
\usepackage[mathscr]{eucal}
\usepackage{color}
\usepackage{booktabs}
\usepackage{lineno}

\large\normalsize

\title{Two timescales in stochastic evolutionary games}
\author{Sabin Lessard$$\footnote{e-mail: sabin.lessard@umontreal.ca}
\\D\'epartement de math\'ematiques et de  statistique\\Universit\'e of Montr\'eal \\
 Montr\'eal H3C 3J7, Canada\\
 }
\date{}
\begin{document}
\maketitle


\section*{Abstract}

Convergence of discrete-time Markov chains with two timescales is a powerful tool to study stochastic evolutionary games in subdivided populations. Focusing on linear games within demes, 
convergence to a continuous-time, continuous-state-space  diffusion process for the strategy frequencies as the unit of time increases to infinity with either the size of the demes or their number yields a strong-migration limit. The same limit is obtained for a linear game in a well-mixed population with effective payoffs that depend on the reproductive values of the demes and identity measures between interacting individuals and competitors within the same demes. The effective game matrix is almost the same under global (hard) selection and local (soft) selection if migration is not too strong.
Moreover, the fixation probability of a strategy introduced as a single mutant is given by a formula 
that can be calculated in the case of a small population-scaled intensity of selection. The first-order effect of selection on this probability, which extends the one-third law of evolution, can also be obtained directly by summing the successive expected changes in the mutant type frequency that involve expected coalescence times of ancestral lines under neutrality. These can be approached by resorting to the existence of two timescales in the genealogical process. On the other hand, keeping the population size fixed but increasing the unit of time to infinity with the inverse of the intensity of migration, convergence to a continuous-time Markov chain after instantaneous initial transitions can be used to obtain a low-migration limit that depends on fixation probabilities within demes in the absence of migration. In the limit of small uniform dispersal, the fixation probability of a mutant strategy in the whole population exceeds its initial frequency if it is risk-dominant over the other strategy with respect to the average payoffs in pairwise interactions in all demes. Finally, introducing recurrent mutation from one strategy to the other, the low-mutation limit takes the inverse of the mutation rate as unit of time. As this rate goes to zero, the average abundance of a strategy in the long run is determined by the fixation probabilities of both strategies when introduced as single mutants in a deme chosen at random.

\noindent \textbf{Keywords and phrases}:   Fixation probability, Diffusion approximation, Average abundance, Coalescent theory, Strong-migration limit, Low-migration limit, Low-mutation limit

\noindent \textbf{Mathematics Subject Classification (2010)}: Primary 91A22 ; Secondary 92D25

\section{Introduction }

Two-timescale arguments have been used for a long time in population genetics. For a finite population in discrete time, the approach has been made rigorous in Ethier and Nagylaki (1980 \cite{*EN1980}) where precise conditions are given for a diffusion approximation to hold in the limit of a large population.  The main idea  is that, if the parameters of the model are appropriately scaled with respect to the population size, some changes in the population structure occur much faster than changes in the allele frequencies. As the population size tends to infinity, there is a separation of timescales so that the former changes occur instantaneously in the timescale of the latter. Moreover, the continuous-time, continuous-state-space limiting process is characterized by its infinitesimal mean and its infinitesimal variance. These are the drift function and diffusion function of the process. They can be used to get expressions for fixation probabilities of given types or moments of the stationary distribution in terms of the population-scaled parameters. The calculations of the expressions, however, can be out of reach unless these parameters are small. 

The existence of two timescales has been applied to several population genetics models in a multi-allelic setting, e.g., to account for genotypic frequencies in Hardy-Weinberg proportions in diploid populations under weak selection (Ethier and Nagylaki, 1980 \cite{*EN1980}) or to give grounds for strong-migration limits in island models as the size or number of the islands tends to infinity (Nagylaki, 1980 \cite{*N1980}, Cherry and Wakeley, 2003 \cite{*CW2003}, Wakeley, 2003 \cite{*W2003}, Wakeley and Takahashi, 2004 \cite{*WT2004}).  See also Slatkin (1981 \cite{*S1981}) for a low-migration limit as the deme size goes to infinity, and Nagylaki (1996 \cite{*N1996}, 1997 \cite{*N1997}) for theoretical studies of infinite populations subdivided into a lattice of colonies under the assumption of weak selection and migration whose intensity is related to the unit of time, and Lessard (2009 \cite{*L2009}) for a unifying approach to kin selection in finite group-structured populations under various assumptions about the timing of dispersal or recolonization of demes that go extinct with fast changes in group type frequencies and slow changes in individual type frequencies.

Separation of timescales has been applied more recently to evolutionary game theory. In Rousset and Billiard (2000 \cite{*RB2000}), for instance, a low-mutation limit and identity measures under neutrality were used to get the first-order effect of selection on the fixation probability of a slightly perturbed strategy in a population of demes homogeneously distributed over a one-dimensional lattice and experiencing localized dispersal. Then, convergence stable strategies (Christiansen, 1991 \cite{*C1991}) can be deduced but with respect to fixation probabilities instead of initial increase in frequency (Rousset, 2003 \cite{*R2003}, Lessard, 2005 \cite{*L2005}). On the other hand, assuming that the individuals occupy the vertices of a finite regular graph and interactions occur between neighbours, Ohtsuki \emph{et al.} (2006 \cite{*OHLN}) used an approximation based on the fact that pair frequencies change quicker than type frequencies under weak selection to show that the benefit-to-cost ratio for altruism must exceed the degree of the graph for the fixation probability of altruism introduced as a single mutant to exceed its initial frequency. This is an important evolutionary property for a strategy to be favored by selection (Nowak \emph{et al.}, 2004 \cite{*NSTF2004}). Another one applied to structured populations (Antal \emph{et al.}, 2008 \cite{*ANT2008}) is to be more abundant on average than the other strategy in the stationary state under recurrent mutation, which is related to fixation probabilities in the limit of low mutation. On the other hand, considering a linear game in an infinite island model with partial dispersal of offspring or local extinction and recolonization of islands or demes, it was shown  in Ohtsuki (2010 \cite{*O2010}) and Lessard (2011a \cite{*L2011a} ) that an effective game matrix  involving identity-by-descent measures, or an inclusive fitness formulation in additive models, could predict the evolutionary dynamics. 

With two possible types of individuals and the possibility for paired individuals to split up from one time step to the next to form new pairs, which is known as the opting-out strategy, it was noted that the changes in pair frequencies occur faster than the changes in type frequencies if selection is weak (Zhang \emph{et al.}, 2016 \cite{*zha82016}; see also K\v{r}ivan and Cressman, 2017 \cite{*KC2017}, for a continuous-time model). Convergence of the stochastic dynamics to a diffusion process in the limit of a large finite population was confirmed in Li and Lessard (2021 \cite{*LL2021}) by checking Ethier and Nagylaki's (1980 \cite{*EN1980}) conditions.
The extension of the result to multiple types is more recondite, however, since one of the main conditions is global convergence of the recurrence system of equations with constant type frequencies in an infinite population. As a matter of fact, with opting-out that depends on the types of the individuals, the equilibrium pair frequencies are generally not in Hardy-Weinberg proportions and their expressions remain to be found, not to mention global convergence to an equilibrium to be ascertained. Such a convergence has been proved, however, for a continuous-time model with given rates for singles of different types to form new pairs and given rates for different pairs to disband (Cressman and K\v{r}ivan, 2022 \cite{*CK2022}). The proof relies on the chemical reaction network theory presented in Feinberg (2019 \cite{*F2019}), and it  is not obvious.

Another important result on Markov chains with two timescales but this time on the same finite state space is due to M\"ohle (1998a \cite{*M1998a}). This result is based on a key lemma about the iterates of transition matrices that extends the well known identity $\lim_{N\rightarrow \infty}(I + A/N)^N=\textrm{e}^A$ for any square matrix $A$. It was first applied to the genealogical process for a sample of genes in a neutral Wright-Fisher diploid population with partial selfing (M\"ohle, 1998a \cite{*M1998a}) and in the case of two separate sexes (M\"ohle, 1998b \cite{*M1998b}). The approach was developed later on to study the strong-migration limit (Notohara, 1993 \cite{*N1993} , Herbots, 1997 \cite{*H1997}, Kroumi and Lessard, 2015 \cite{*KL2015}) and the  island model with partial dispersal (Wakeley, 1998 \cite{*W1998}, Nordborg, 2001 \cite{*N2001}, Lessard and Wakeley, 2004 \cite{*LW2004} , Wakeley and Lessard, 2006 \cite{*WL2006}). It was used to get the first-order effect of selection on the fixation probability in a well-mixed population and in an island model even when a diffusion approximation does not hold as a result of highly skewed distribution of family size (Lessard and Ladret, 2007 \cite{*LL2007}, Lessard, 2011b \cite{*L2011b} ) or when viabilities or payoffs in pairwise interactions are random variables as a consequence of stochastic fluctuations in the environment (Kroumi and Lessard, 2024a \cite{*KL2024a}, 2024b \cite{*KL2024b}). A direct analysis of a low-migration limit in a Moran type population of fixed finite size was performed recently (Pires and Broom, 2024 \cite{*PB2024} ).

In this paper, we will review some of the main results in evolutionary game theory that involve two-timescale arguments, providing simplified proofs and some new insight. We will focus on results based on Ethier and Nagylaki's (1980 \cite{*EN1980}) conditions for a discrete-time Markov chain with two timescales to converge to a diffusion approximation in the limit of a large population, and  M\"ohle's (1998a \cite{*M1998a}) lemma on the convergence to a continuous-time Markov chain on the same finite state space. For completeness, these tools will be recalled in Appendices A and B, as well as the backward-time genealogical processes that will come into play in their applications. The objective is to provide some new insight on two timescale arguments for evolutionary games in subdivided populations.

Throughout the paper, we consider two strategies, $A$ and $B$, in a population subdivided into a finite number of finite demes. Time is discrete and the reproductive success of the strategies from one time step to the next is determined by the payoffs received in linear games within demes. The details of the model are presented in Section 2. In Section 3, 
we deduce the strong-migration limit as the deme size goes to infinity under 
a general migration pattern of offspring after selection and an updating rule within demes according to the Wright-Fisher model or the Moran model. In Section 4, the demes are distinguished only by their composition and the strong-migration limit is obtained under partial uniform dispersal of offspring as the number of demes goes to infinity. In Section 5, we calculate the first-order effect of selection on the fixation probability of $A$ introduced as a single mutant by using either a diffusion approximation for the frequency of $A$ or a direct approach based on the successive expected changes in the frequency of $A$ that involve expected coalescence times of ancestral lines under neutrality. In Section 6, the low-migration limit of the model in Section 2 is deduced  
by taking the inverse of the intensity of migration to other demes as unit of time as this intensity goes to $0$. In Section 7, it is the low-mutation limit that is deduced  by introducing a recurrent mutation probability taken as unit of time as this probability goes to $0$. Finally, in Section 8, we draw some conclusions on the strengths and weaknesses of diffusion approximations and direct approaches based on convergence of transition matrices  to deal with two timescales in evolutionary game theory.

\section{Model}

We consider a population subdivided into $D$ demes or colonies with $N$ resident individuals in deme $i$ for $i=1, \ldots, D$ as in Karlin (1982 \cite{*K1982}) to study selection-migration structures in population genetics but with haploid individuals. Moreover, the deme size is assumed  finite as in Nagylaki (1980 \cite{*N1980}), and assumed to be the same for all demes for simplicity. There are two types of individuals, $A$ and $B$, and the frequency of $A$ in deme $i$ is denoted by $x_i$ for $i=1, \ldots, D$.  The individuals produce locally large numbers of offspring in equal proportions. 
Following random pairwise interactions between offspring within demes, the average payoffs to $A$ and $B$ in deme $i$ are given by 
\begin{align}
w_A(i, x_i)=a_i x_i + b_i (1-x_i)
\end{align}
and 
\begin{align}
w_B(i, x_i)=c_i x_i + d_i (1-x_i),
\end{align}
respectively, for some payoffs $a_i, b_i, c_i, d_i$ for $i=1, \ldots, D$.  Assuming that the corresponding viabilities of offspring are given by $1+sw_A(i, x_i)$ and $1+sw_B(i, x_i)$, respectively,
where $s>0$ represents some small intensity of selection, the frequency of $A$ among the viable offspring in deme $i$ is 
\begin{align}
\tilde{x}_i=\frac{x_i(1+sw_A(i, x_i))}{1+s\bar{w}(i, x_i)}\approx x_i + s x_i(1-x_i)(w_A(i, x_i)-w_B(i, x_i)),
\end{align}
where $\bar{w}(i, x_i)=x_iw_A(i, x_i) +(1-x_i)w_B(i, x_i)$ is the average payoff in deme $i$ and 
\begin{align}
w_A(i, x_i)-w_B(i, x_i)= (a_i-c_i)x_i + (b_i-d_i)(1-x_i),
\end{align}
 for $i=1, \ldots, D$. 

Viability selection is followed by migration of viable offspring. We let $\mu_{ij}$ be the probability for an offspring in deme $i$ to move to deme $j$, for $i, j=1, \ldots, D$. 
Under local selection (or soft selection, see Christiansen, 1975 \cite{*C1975}) so that the contributions of the demes in viable offspring are all the same, the probability for an offspring in deme $i$ to come from deme $j$ is given by 
\begin{align}
m_{ij}^{}=\frac{\mu_{ji}}{\sum_{k=1}^D \mu_{ki}}
\end{align}
for $i,j=1, \ldots, D$. The stochastic matrix $\mathbf{M}^{}=(m_{ij}^{})_{i,j=1}^D$ is the backward migration matrix under soft selection. This matrix does not depend on selection. Under global selection (or hard selection, see Christiansen, 1975 \cite{*C1975}) so that  the contribution of a deme in viable offspring is proportional to the average viability in the deme, however, the above  probability becomes \begin{align}
m_{ij}^{*}=\frac{(1+s\bar{w }(i, x_i))\mu_{ji}}{\sum_{k=1}^D (1+s\bar{w}(k, x_k))\mu_{ki}}\approx m_{ij}\left(1+ s \sum_{k=1}^D (\bar{w}(i, x_i)-\bar{w }(k, x_k))m_{ik} \right)
\end{align}
for $i,j=1, \ldots, D$.  
The backward migration matrix under hard selection $\mathbf{M}^{*}=(m_{ij}^{*})_{i,j=1}^D$ is generally different from $\mathbf{M}$ unless there is no selection ($s=0$). 

It will be assumed throughout that the backward migration matrix under neutrality, which corresponds to $\mathbf{M}$, 
is irreducible and aperiodic. Therefore, the ergodic theorem for Markov chains (see, e.g., Karlin and Taylor, 1975 \cite{*KT1975}) guarantees that 
\begin{align}
\lim_{\tau \rightarrow \infty}\mathbf{M}^{\tau} =\mathbf{1} \boldsymbol{\pi}^T,
\end{align}
where $\mathbf{1}$ denotes a column vector with all components equal to $1$, while $\boldsymbol{\pi}^T=(\pi_1, \ldots, \pi_D)$ with $T$ for transpose is a positive row vector satisfying $\boldsymbol{\pi}^T\mathbf{M}=\boldsymbol{\pi}^T$ and $\sum_{i=1}^D \pi_i=1$, which defines a stationary probability distribution. Note that the probability $\pi_i$ represents the fraction of time steps back that the ancestral line of an individual chosen at random would spend in deme $i$ in the long run in the absence of selection, for $i=1, \ldots, D$. This probability corresponds to the contribution of deme $i$ to all future generations in a neutral model, known as  a reproductive value (Fisher, 1930 \cite{*F1930}).

After soft selection and migration of offspring, the frequency of $A$ among the offspring in deme $i$ is 
\begin{align}
\tilde{\tilde{x}}_i=\sum_{j=1}^D m_{ij} \tilde{x}_j
\end{align}
for $i=1, \ldots, D$. The life cycle is completed by random sampling of offspring within demes to replace resident individuals. If all individuals in all demes are replaced at a time, then we have a Wright-Fisher updating rule (Fisher, 1930 \cite{*F1930}, Wright, 1931 \cite{*W1931}; see Ewens, 2004 \cite{*E2004}). Then, the new frequency of $A$ in deme $i$, denoted by $x^{\prime}_i$, is such that $Nx^{\prime}_i$ follows a binomial probability distribution whose parameters are $N$ and $\tilde{\tilde{x}}_i$, for $i=1, \ldots, D$. On the other hand, if only one individual chosen at random in a deme chosen at random is replaced, then we have a Moran type model (Moran, 1958 \cite{*M1958}). In this case, the individual replaced is in deme $i$ with probability $1/D$, and then it is replaced by an $A$ individual  with probability $\tilde{\tilde{x}}_i$ or otherwise by a $B$ individual, for $i=1, \ldots, D$. These updating rules will change the weighted frequency of $A$ in the whole population defined as
\begin{align}
x^{}=\sum_{i=1}^D \pi_i x_i,
\end{align}
where $(\pi_1, \ldots, \pi_D)$ is the stationary distribution associated with the  backward migration matrix $\mathbf{M}$.

\section{Large-demes limit}

In the Wright-Fisher model, the weighted frequency of $A$ in the whole population at the next time step is
\begin{align}
x^{\prime}=\sum_{i=1}^D \pi_i x^{\prime}_i.
\end{align}
Then, given a population state $\mathbf{x}^T=(x_1, \ldots, x_D)$,  the change in  the weighted frequency of $A$, denoted by $\Delta x = x^{\prime}-x$, has conditional expectation
\begin{align}\label{conditionalexpectation}
\mathbb{E}(\Delta x \, | \, \mathbf{x})=\sum_{i=1}^D \pi_i(\tilde{\tilde{x}}_i-x_i) 
\end{align}
and conditional variance
\begin{align}\label{conditionalvariance}
\mathbb{V}(\Delta x \, | \, \mathbf{x})=\frac{1}{N}\sum_{i=1}^D \pi_i^2\tilde{\tilde{x}}_i(1- \tilde{\tilde{x}}_i).
\end{align}
Assuming soft selection with an intensity of selection $s=\sigma/N$, we have the approximations
\begin{align}\label{appro1}
\mathbb{E}(\Delta x \, | \, \mathbf{x})= \frac{\sigma}{N}\sum_{j=1}^D \pi_j x_j(1-x_j)(w_A(j, x_j)-w_B(j, x_j))+o(1/N)
\end{align}
and 
\begin{align}\label{appro2}
\mathbb{V}(\Delta x \, | \, \mathbf{x})= \frac{1}{N}\sum_{i=1}^D \pi_i^2 \left( \sum_{j=1}^D m_{ij}x_j\right)\left(1- \sum_{j=1}^D m_{ij}x_j\right) +o(1/N).
\end{align}
It can also be checked that $\mathbb{E}((\Delta x)^4 \, | \, \mathbf{x})=o(1)$, while $y_i=x_i-x$ verifies
\begin{align}
\mathbb{E}(y^{\prime}_i-y_i) \, | \, \mathbf{x})=\tilde{\tilde{x}}_i-x_i- \mathbb{E}(\Delta x \, | \, \mathbf{x})= \sum_{j=1}^D m_{ij}y_j-y_i + o(1)
\end{align}
and $\mathbb{V}(y^{\prime}_i-y_i \, | \, \mathbf{x})=o(1)$ for $i=1, \ldots, D$. Moreover, the zero solution of the system of recurrence equations
\begin{align}
Y^{\prime}_i= \sum_{j=1}^D m_{ij}Y_j
\end{align}
for $i=1, \ldots, D$ under the constraint $\sum_{i=1}^D \pi_i Y_i=0$ is globally asymptotically stable. As a matter of fact, the $\tau$-th iterate $\mathbf{Y}^{(\tau)}=(Y^{(\tau)}_1, \ldots, Y^{(\tau)}_D)^T$ verifies 
\begin{align}
\mathbf{Y}^{(\tau)}=\mathbf{M}^{\tau}\mathbf{Y}^{(0)} \rightarrow \mathbf{1} \boldsymbol{\pi}^T \mathbf{Y}^{(0)}=\mathbf{0}
\end{align}
as $\tau \rightarrow \infty$ owing to the ergodic theorem and the constraint on $\mathbf{Y}^{(0)}$, where $\mathbf{0}$ denotes a column vector with all $0$ components.

Applying a convergence result for Markov chains with two timescales (Ethier and Nagylaki, 1980 \cite{*EN1980}, see Appendix A) as in  Nagylaki (1980 \cite{*N1980}), the weighted frequency of $A$ in the limit of a large deme size $N$ with this number of life cycles as unit of time is described by a continuous-time, continuous-state-space diffusion process 
whose infinitesimal mean is
  \begin{align}
m(x) 
&= \sigma x(1-x)\sum_{j=1}^D \pi_j (w_A(j, x)-w_B(j, x))\nonumber\\
&=\sigma x(1-x)\left( x\sum_{j=1}^D \pi_j (a_j-c_j) +(1-x) \sum_{j=1}^D \pi_j (b_j-d_j)\right),
\end{align}
and infinitesimal variance 
\begin{align}
v(x)
= x(1-x)\sum_{i=1}^D \pi_i^2.
\end{align}
These expressions are obtained from Eqs. (\ref{appro1}) and (\ref{appro2}) for $x_j=x$ for $j=1, \ldots, D$. 

It is worth noting that the same infinitesimal mean is obtained if there is only one deme and the payoffs 
to $A$ and $B$ in random pairwise interactions are given by  the entries of the matrix
\begin{equation}\label{matrix1}
\begin{pmatrix}
\sum_{j=1}^D \pi_j a_j&\sum_{j=1}^D \pi_j b_j\\
\sum_{j=1}^D \pi_j c_j&\sum_{j=1}^D \pi_j d_j \\
\end{pmatrix},
\end{equation}
called an effective game matrix.

In the Moran model under soft selection with an intensity of selection $s=2\sigma/N$, the change in  the weighted frequency of $A$ has conditional expectation
\begin{align}\label{appro5}
\mathbb{E}(\Delta x \, | \, \mathbf{x})&=\frac{1}{DN}\sum_{i=1}^D \pi_i((1-x_i) \tilde{\tilde{x}}_i- x_i(1- \tilde{\tilde{x}}_i)\nonumber\\
&=\frac{1}{DN}\sum_{i=1}^D \pi_i( \tilde{\tilde{x}}_i- x_i)\nonumber\\
&= \frac{2\sigma}{DN^2}\sum_{j=1}^D \pi_j x_j(1-x_j)(w_A(j, x_j)-w_B(j, x_j))+o(1/N^2)
\end{align}
and conditional variance
\begin{align}
\mathbb{V}(\Delta x \, | \, \mathbf{x})&=\mathbb{E}((\Delta x)^2 \, | \, \mathbf{x}) +o(1/N^2)\nonumber\\
&=\frac{1}{DN^2}\sum_{i=1}^D \pi_i^2((1-x_i) \tilde{\tilde{x}}_i+ x_i(1- \tilde{\tilde{x}}_i))+o(1/N^2)\nonumber\\
&= \frac{1}{DN^2}\sum_{i=1}^D \pi_i^2 \left(x_i+(1-2x_i) \sum_{j=1}^D m_{ij}x_j\right) +o(1/N^2).
\end{align}
All the other conditions for a diffusion approximation in the limit of a large deme size $N$ with $DN^2/2$ time steps as unit of time are satisfied. Moreover, the infinitesimal mean and variance of the limiting diffusion are the same as previously. 

It is worth noting that, if viability selection occurs after migration, then $x_j$ is replaced by $\sum_{k=1}^D m_{jk}x_k$ for $j=1, \ldots, D$ in the conditional expected changes in the weighted frequency of $A$ given in Eqs. (\ref{appro1}) and (\ref{appro5}), which leads to the same diffusion approximation.

\section{Many-demes limit}

In this section, we consider a large number $D$ of groups or demes of fixed size $N$ which differ only by their composition as in Wakeley (2003 \cite{*W2003}). Moreover, we assume partial uniform dispersal of offspring so that a proportion $1-m$ of offspring stay in their natal deme while a proportion $m$ migrate to a deme chosen at random among all demes including the natal deme. We let $z_k$ be the frequency of demes containing $k$ individuals of type $A$ and $N-k$ individuals of type $B$ for a frequency of $A$ given by $x(k)=k/N$. In such a deme, the average payoffs to $A$ and $B$ offspring are represented by \begin{align}
w_A(x(k))=ax(k) + b(1-x(k))
\end{align} 
and 
\begin{align}
w_B(x(k))=cx(k) + d (1-x(k)),
\end{align} 
respectively. These average payoffs bring the frequency of $A$ among the offspring in the deme to
\begin{align}
\tilde{x}(k)=\frac{x(k)(1+sw_A(x(k)))}{1+s\bar{w}(x(k))}\approx x(k)+ s x(k)\left(1-x(k)\right)(w_A(x(k))-w_B(x(k)))
\end{align}
with $\bar{w}(x(k))=x(k)w_A(x(k)) +(1-x(k))w_B(x(k))$ and 
\begin{align}
w_A(x(k))-w_B(x(k))=(a-c)x(k)+ (b-d)(1-x(k))
\end{align}
 for $k=0,1, \ldots, N$. Assuming soft selection with intensity of selection $s=\sigma/(ND)$, uniform dispersal of offspring with probability $m$ and sampling of $N$ offspring in each deme according to a Wright-Fisher model to start the next generation,  the frequency of $A$ that was $x(k)=k/N$ in a deme becomes $x(l)=l/N$, independently of what happens in the other demes, with probability given by
\begin{align}
P_{k,l}= 
{N\choose l}\tilde{\tilde{x}}(k)^l (1-\tilde{\tilde{x}}(k))^{N-l},
\end{align}
where
\begin{align}\label{soft}
\tilde{\tilde{x}}(k)&=(1-m)\tilde{x}(k)+m\sum_{l=0}^N z_l\tilde{x}(l)\nonumber\\
&\approx (1-m)x(k)+mx +(1-m)\frac{\sigma}{ND}x(k)\left(1-x(k)\right)(w_A(x(k))-w_B(x(k))) \nonumber\\
&\quad\quad\quad\quad\quad\quad\quad+m \frac{\sigma}{ND}\sum_{l=0}^Nz_lx(l)\left(1-x(l)\right)(w_A(x(l))-w_B(x(l))) +o(1/D).
\end{align}
Using Eqs. (\ref{conditionalexpectation}) and (\ref{conditionalvariance}) with $\pi_i=1/D$ for $i=1, \ldots, D$,  the frequency of $A$ in the whole population represented by $x= \sum_{k=0}^N z_k x(k)$ has a change $\Delta x= x^{\prime}-x$ whose conditional expectation is
\begin{align}\label{appro10}
\mathbb{E}(\Delta x \, | \, \mathbf{z})&=\sum_{k=0}^N z_k\left(\tilde{\tilde{x}}(k)-x(k)\right) \nonumber\\
&= \frac{\sigma}{ND}\sum_{k=0}^N z_k x(k)(1-x(k))(w_A(x(k))-w_B(x(k)))+o(1/D),
\end{align}
and conditional variance
\begin{align}\label{}
\mathbb{V}(\Delta x \, | \, \mathbf{z})&=\frac{1}{ND}\sum_{k=0}^N z_k\tilde{\tilde{x}}(k)(1- \tilde{\tilde{x}}(k))\nonumber\\
&=\frac{1}{ND}\sum_{k=0}^N z_k\left( (1-m)x(k)+mx\right)\left(1-  (1-m)x(k)-mx\right) + o(1/D),
\end{align}
given a current population state $\mathbf{z}=(z_0, z_1, \ldots, z_N)^T$. 

Now, let $\mathbf{v}(x)^T=(v_0(x), v_1(x), \ldots, v_N(x))$ be  the positive frequency vector that satisfies the equation
$\mathbf{v}(x)^T=\mathbf{v}(x)^T\mathbf{P}^{*}$, where the entries of $\mathbf{P}^{*}=\left(P^{*}_{k,l}\right)_{k,l=0}^N$ are given by 
\begin{align}
P^{*}_{k,l}= 
{N\choose l}\left((1-m)x(k)+mx \right)^l \left(1-(1-m)x(k)-mx\right)^{N-l},
\end{align}
for $k, l=0, 1, \ldots, N$. This is the stationary distribution for the deme types in the population when there is an infinite number of demes in the absence of selection with a frequency of $A$ given by $x$. Now, if we define $y_l=z_l - v_l(x)$ for $l=0,1, \ldots, N$, where $x=\sum_{l=0}^N z_l x(l)$, we have 
\begin{align}
\mathbb{E}(y^{\prime}_l-y_l \, | \, \mathbf{z})= \sum_{k=0}^N z_kP^{}_{k,l} - \mathbb{E}(v_l(x^{\prime})\, | \, \mathbf{z} )- y_l=\sum_{k=0}^N y_kP^{*}_{k,l} +v_l(x)- \mathbb{E}(v_l(x^{\prime}) \, | \, \mathbf{z} )- y_l
\end{align}
for $l=0,1, \ldots, N$. 

We will make use of an important property conjectured in Wakeley (2003 \cite{*W2003}) and shown in
Lessard (2007 \cite{*L2007}) (see also Eq. (7.7) in Lessard, 2009 \cite{*L2009}). It says that there exist constants $r_k$ for $k=1, \ldots, N$ such that
\begin{align}
v_l (x^{\prime}) = v_l(x) + \sum_{k=1}^N r_k (x^{\prime} - x)^k
\end{align}
for $l=0,1, \ldots, N$. Using the facts that $x$ is bounded by $1$ and the first two conditional moments of $\Delta x=x^{\prime} - x$ are of order $o(1)$, we have 
\begin{align}
 \mathbb{E}(v_l(x^{\prime})\, | \, \mathbf{z} )=v_l(x) +o(1),
 \end{align} 
 from which
 \begin{align}
\mathbb{E}(y^{\prime}_l-y_l \, | \, \mathbf{z})=\sum_{k=0}^N y_kP^{*}_{k,l} - y_l+o(1)
\end{align}
for $l=0,1, \ldots, N$.
Moreover, the zero solution of the system of recurrence equations
\begin{align}
Y^{\prime}_l= \sum_{k=0}^N Y_kP^{*}_{k,l} 
\end{align}
for $l=0,1,  \ldots, N$ under the constraint $\sum_{k=0}^N Y_k=0$ is globally asymptotically stable. As a matter of fact, the $\tau$-th iterate $\mathbf{Y}^{(\tau)T}=(Y^{(\tau)}_0, Y^{(\tau)}_1,\ldots, Y^{(\tau)}_N)$ verifies 
\begin{align}
\mathbf{Y}^{(\tau)T}=\mathbf{Y}^{(0)T}\mathbf{P}^{*\tau} \rightarrow  \mathbf{Y}^{(0)T} \mathbf{1} \boldsymbol{v}(x)^T =\mathbf{0}
\end{align}
as $\tau \rightarrow \infty$ owing to the ergodic theorem and the constraint on $\mathbf{Y}^{(0)}$. It can also be checked that $\mathbb{E}((\Delta x)^4 \, | \, \mathbf{z})=o(1)$ and $\mathbb{V}(y^{\prime}_l-y_l \, | \, \mathbf{z})=o(1)$ for $l=0,1, \ldots, N$ (see Lessard, 2009 \cite{*L2009}), Kroumi and Lessard, 2024a \cite{*KL2024a}, 2024b \cite{*KL2024b}).

Applying again Ethier and Nagylaki's (1980 \cite{*EN1980})) result in the limit of a large number $D$ of demes of constant size $N$ with $ND$ time steps as unit of time, the frequency of $A$ is described by a diffusion process whose infinitesimal mean is
  \begin{align}\label{infmean0}
m(x) &=  \sigma \sum_{k=0}^N v_k(x) x(k)(1-x(k))(w_A(x(k))-w_B(x(k)))\nonumber\\
&=\sigma \left((a-c) \sum_{k=0}^N v_k(x) x(k)^2(1-x(k)) + (b-d) \sum_{k=0}^N v_k(x) x(k)(1-x(k))^2\right)
\end{align}
and infinitesimal variance 
\begin{align}\label{}
v(x)&=\sum_{k=0}^N v_k(x)\left( (1-m)x(k)+mx\right)\left(1-  (1-m)x(k)-mx\right).
\end{align}
This infinitesimal variance is the probability for two offspring chosen at random in the same deme \emph{after dispersal} to be of types $A$ and $B$ in this order in a stationary population of an infinite number of demes in the absence of selection in which the frequency of $A$ is fixed and equal to $x$. This is the probability for the two offspring to have two ancestors in different demes in such a population, represented by $\tilde{f}_{22}$, times the frequencies of $A$ and $B$. Therefore, we have
\begin{align}\label{}
v(x)&=\tilde{f}_{22}x(1-x).
\end{align}
As for the infinitesimal mean, it takes the form
   \begin{align}\label{infmean5}
m(x) 
=\sigma x(1-x)\left((a-c) \left(xf_{33} + f_{32}/3 \right)+ (b-d) \left((1-x)f_{33} + f_{32}/3 \right)\right),
\end{align}
where $f_{32}$ and $f_{33}$ designate  the probabilities for  three offspring chosen at random in the same deme \emph{before dispersal} to have two and three ancestors in different demes, respectively, under the above assumptions. These differ little from the corresponding probabilities for three offspring chosen at random in the same deme after dispersal, denoted by $\tilde{f}_{32}$ and $\tilde{f}_{33}$, respectively. See Appendix E or Rousset (2002 \cite{*R2002}) for exact expressions of $\tilde{f}_{22}$, $f_{32}$ and $f_{33}$. 
For $m$ small and $N$ large, we have the approximations
\begin{subequations}\label{}
\begin{align}\label{}
 &\tilde{f}_{22}\approx \frac{2\nu}{1+2\nu},\\
 &f_{32}\approx \frac{3\nu}{(1+\nu)(1+2\nu)},\\
  &f_{33}\approx \frac{2\nu^2}{(1+\nu)(1+2\nu)},
  \end{align}
  \end{subequations}\label{}
 where $\nu=Nm$ is the deme-scaled dispersal rate. Note that $1-\tilde{f}_{22}$ corresponds to Wright's (1931 \cite{*W1931}) fixation index $F_{ST}$.
 
It is easy to check that the effective game matrix 
 \begin{align}\label{matrix2}
f_{33}\begin{pmatrix}
a&b\\
c&d \\
\end{pmatrix}+ \frac{f_{32}}{3}\begin{pmatrix}
a&a+b-c\\
d+c-b&d \\
\end{pmatrix}
\end{align}
for random pairwise interactions in a well-mixed population gives
the same infinitesimal mean. Moreover, using the relationships $f_{22}=1-f_{21}$, $f_{32}=3(f_{21}-f_{31})$ and $f_{33}=1-f_{31}-f_{32}$, this effective game matrix can be expressed as
 \begin{align}\label{matrix2bis}
f_{22}\begin{pmatrix}
a&b\\
c&d \\
\end{pmatrix}+ f_{21}\begin{pmatrix}
a&a\\
d&d \\
\end{pmatrix}
 -f_{31}\begin{pmatrix}
a&a\\
d&d \\
\end{pmatrix}- \frac{f_{32}}{3}\begin{pmatrix}
a&b\\
c&d \\
\end{pmatrix}- \frac{f_{32}}{3}\begin{pmatrix}
a&c\\
b&d \\
\end{pmatrix}.
\end{align}
See the Discussion section for an interpretation.

In the case of selection after dispersal, the frequency $x(k)$ is replaced by $(1-m)x(k)+mx$ in Eqs. (\ref{appro10}) and (\ref{infmean0}) for $k=0, 1, \ldots, N$, and then  the probabilities $f_{22}, f_{21}, f_{33}, f_{32}$ and $f_{31}$ for offspring before dispersal in Eqs. (\ref{infmean5}), (\ref{matrix2}) and (\ref{matrix2bis}) are replaced by the corresponding probabilities for offspring after dispersal $\tilde{f}_{22}, \tilde{f}_{21}, \tilde{f}_{33}, \tilde{f}_{32}$ and $ \tilde{f}_{31}$.

In the case of hard selection before dispersal,  the expression in Eq. (\ref{soft}) for the frequency of $A$ in a deme after selection and dispersal is replaced by
\begin{align}\label{hard}
\tilde{\tilde{x}}(k)&=\frac{(1-m)\tilde{x}(k)\bar{w}(x(k))+m\sum_{l=0}^N z_l\tilde{x}(l)\bar{w}(x(l))}{(1-m)\bar{w}(x(k))+m\bar{w}},
\end{align}
where
\begin{align}\label{}
\bar{w}=\sum_{l=0}^N z_l\bar{w}(x(l))
\end{align}
is the mean payoff in the whole population.
The approximation of this frequency differs from the one given in Eq. (\ref{soft}) by the terms
\begin{align}\label{}
&\frac{\sigma}{ND}(1-m)x(k)\bar{w}(x(k)) + \frac{\sigma}{ND}m\sum_{l=0}^N z_l\tilde{x}(l)\bar{w}(x(l))\nonumber\\
&-\frac{\sigma}{ND} \left((1-m)x(k)+mx  \right) \left((1-m)\bar{w}(x(k)) +m\bar{w } \right). 
\end{align}
This leads to an infinitesimal mean with the extra term
  \begin{align}
&\sigma m(2-m) \sum_{k=0}^N v_k(x) x(k)(\bar{w}(x(k))-\bar{w})\nonumber\\
&= \sigma m(2-m) \left( (a-c)\left(f_{31}+\frac{2f_{32}}{3}x\right) + (b-d)\left( f_{31}+\frac{2f_{32}}{3}(1-x)\right) +(c-b)f_{21}\right).
\end{align}
Then, the corresponding effective game matrix in a well-mixed population has the extra term
 \begin{equation}\label{matrix3}
m(2-m)\left[\frac{2f_{32}}{3}\begin{pmatrix}
a&b\\
c&d \\
\end{pmatrix}+ f_{31}\begin{pmatrix}
a&a+b-c\\
d+c-b&d \\
\end{pmatrix}
+ f_{21}\begin{pmatrix}
c&c\\
b&b \\
\end{pmatrix}\right].
\end{equation}
This yields an effective game matrix that can be expressed as
 \begin{align}\label{matrix3}
f_{22}\begin{pmatrix}
a&b\\
c&d \\
\end{pmatrix}+ f_{21}\begin{pmatrix}
a&a\\
d&d \\
\end{pmatrix}
 -(1-m)^2\left(f_{31}\begin{pmatrix}
a&a\\
d&d \\
\end{pmatrix}+ \frac{f_{32}}{3}\begin{pmatrix}
a&b\\
c&d \\
\end{pmatrix}+ \frac{f_{32}}{3}\begin{pmatrix}
a&c\\
b&d \\
\end{pmatrix}\right).
\end{align}
See the Discussion section for an interpretation of this result.

\section{Fixation probability}

In this section, we will be interested in conditions for selection to favour the evolution of $A$, more precisely in conditions for the fixation probability of $A$ introduced as a single mutant to exceed what it would be under neutrality (Nowak \emph{et al.}, 2004 \cite{*NSTF2004}). These conditions can be used to identify convergence stable strategies (Christiansen, 1991 \cite{*C1991}) in finite populations with respect to fixation probabilities instead of initial increase in frequency (Rousset, 2003 \cite{*R2003}, Lessard, 2005 \cite{*L2005}).

When the frequency of $A$ over discrete time steps, represented by $(x(\tau))_{\tau\geq o}$, converges to a diffusion process whose infinitesimal mean and variance are $m(x)$ and $v(x)$ for $x \in [0, 1]$, respectively, then the fixation probability of $A$ can be approximated as

  \begin{align}
\rho_A \approx \frac{ \int_{0}^{x(0)} \psi (y) dy}{\int_{0}^{1} \psi (y) dy},
\end{align}
where $x(0)$ is the initial frequency of $A$ and
  \begin{align}
\psi (y)= \exp \left(-2 \int_0^y \frac{m(x)}{v(x)}dx \right)
\end{align}
for $y \in [0, 1]$ (see, e.g., Ewens, 2004 \cite{*E2004}). If
  \begin{align}
\frac{m(x)}{v(x)}= \sigma \sum_{k=0}^n a_k x^k,
\end{align}
for some coefficients $a_0, a_1, \ldots, a_k$ and small parameter $\sigma >0$, then we have
  \begin{align}
\rho_A
&\approx x(0)\left(1 + 2\sigma\int_0^1 \int_0^y \left(  \sum_{k=0}^n a_k x^k\right)dxdy\right)\nonumber\\
&= x(0) \left(1 + 2\sigma \sum_{k=0}^n a_k\int_0^1  (1-x) x^kdx\right)\nonumber\\
&=x(0)\left(1 + 2\sigma\sum_{k=0}^n \frac{a_k}{(k+1)(k+2)}\right).
\end{align}
This approximation can be used for the fixation probability of $A$ in Sections 3 and 4 if 
 the  population-scaled  intensity of selection $\sigma$ is small enough, which means weak selection, actually selection weaker than random drift.

Another way to get the first-order approximation of the fixation probability of $A$ with respect to the intensity of selection is to consider the successive conditional expected changes in the frequency of $A$ as in Rousset (2003 \cite{*R2003}) and Lessard and Ladret (2007 \cite{*LL2007}). As a matter of fact, the fixation probability of $A$ is the expected value of the frequency of $A$ at time step $\tau_0$ as $\tau_0 \rightarrow \infty$. This frequency can be written as
  \begin{align}
x(\tau_0) = x(0) +\sum_{\tau=0}^{\tau_0} \Delta x(\tau), 
\end{align}
where $\Delta x(\tau) =x(\tau +1) - x(\tau)$ for $\tau =0, 1, \ldots, \tau_0$. Properties of expectation and conditional expectation lead to
  \begin{align}
\rho_A= \lim_{\tau_0 \rightarrow \infty} \mathbb{E}\left(x(\tau_0) \right)= \mathbb{E}\left(x(0)\right)+\sum_{\tau=0}^{\infty} \mathbb{E}\left(\Delta x(\tau)\right),
\end{align}
where 
 \begin{align}
 \mathbb{E}\left(\Delta x(\tau)\right) =  \mathbb{E}\left( \mathbb{E}\left(\Delta x(\tau)| \mathbf{x}(\tau)\right)\right)\approx \mathbb{E}_0\left( \mathbb{E}\left(\Delta x(\tau)| \mathbf{x}(\tau)\right)\right)
\end{align}
and $ \mathbf{x}(\tau)$ stands for the population state  at time step $\tau \geq 0$, while $\mathbb{E}_0$ denotes expectation in the absence of selection. 

For the weighted frequency of $A$ in Section 3 under the assumptions soft selection and updating according to either the Wright-Fisher model or the Moran model, Eqs. (\ref{appro1}) and (\ref{appro5}) imply
 \begin{align}
 \mathbb{E}\left(\Delta x(\tau)| \mathbf{x}(\tau)\right)=s\sum_{j=1}^D \pi_jx_j(\tau) (1-x_j(\tau))((a_j-c_j) x_j(\tau) + (b_j-d_j) (1-x_j(\tau))) + o(s),
\end{align}
where $s$ is the intensity of selection and $x_j(\tau)$ is the frequency of $A$ in deme $j$ at time step $\tau \geq 0$ for $j=1, \ldots, D$. Then, the fixation probability of $A$ introduced as a single mutant in deme $i$ so that the initial weighted frequency of $A$ is $x(0)=\pi_i/N$ can be approximated as $ \pi_i/N$ plus the first-order effect of selection given by
  \begin{align}
 s\left(\sum_{j=1}^D \pi_j(a_j-c_j) \mathbb{E}_0\left(\sum_{\tau=0}^{\infty} x_j(\tau)^2 (1-x_j(\tau))\right)+\sum_{j=1}^D \pi_j(b_j-d_j) \mathbb{E}_0\left(\sum_{\tau=0}^{\infty} x_j(\tau) (1-x_j(\tau))^2\right)\right),
\end{align}
where $\mathbb{E}_0$ denotes expectation in a neutral model ($s=0$). Looking at the genealogical process of three offspring chosen at random in deme $j$ before dispersal (see Appendices C and D) and defining $S_3$ and $S_2$ as the number of time steps back with $3$ and $2$ ancestors, respectively, we have the approximation
 \begin{align}
\mathbb{E}_0\left(\sum_{\tau=0}^{\infty} x_j(\tau)^2 (1-x_j(\tau))\right)
&\approx\sum_{\tau=0}^{\infty} (1/3)\mathbb{P}_0(S_3 + S_2 > \tau\geq S_3)\nonumber\\
&\quad\quad\quad\quad \times \mathbb{P}_0(\textrm{ancestors }  AB \textrm{ in this order at time step $0$} )\nonumber\\
&\approx\frac{\pi_i}{3N} (\mathbb{E}_0(S_3 +S_2)-\mathbb{E}_0(S_3))\nonumber\\
&=\frac{\pi_i}{3N}\mathbb{E}_0(S_2).
\end{align}
This is for $N$ large enough so that multiple coalescence events of ancestral lines can be ignored. Then, $1/3$ is the probability for $2$ particular lines out of $3$ to coalesce given that a coalescence event occurs. Similarly, we have
  \begin{align}
\mathbb{E}_0\left(\sum_{\tau=0}^{\infty} x_j(\tau) (1-x_j(\tau))^2\right)&\approx \sum_{\tau=0}^{\infty} \mathbb{P}_0(S_3 > \tau) \mathbb{P}_0(\textrm{ancestors }  ABB \textrm{ in this order at time step $0$} )
\nonumber\\
&\quad+\sum_{\tau=0}^{\infty}(1/3) \mathbb{P}_0(S_3 + S_2 > \tau \geq S_3) \nonumber\\
&\quad\quad\quad\quad\quad \times \mathbb{P}_0(\textrm{ancestors }  AB \textrm{ in this order at time step $0$} )\nonumber\\
&\approx\frac{\pi_i}{N} \mathbb{E}_0(S_3 ) +\frac{\pi_i}{3N} \mathbb{E}_0(S_2 )\nonumber\\
&\approx\frac{2\pi_i}{3N} \mathbb{E}_0(S_2 ).
\end{align}
Here, we use the fact that, with an appropriate time scale as $N \rightarrow \infty$, $S_2$ and $S_3$ in the neutral model tend to exponential random variables with parameters $1$ and $3$, respectively (see Appendices C and D).  This implies that $\mathbb{E}_0(S_2 )\approx 3\mathbb{E}_0(S_3 )$.

 We conclude that $\rho_A>\pi_i/N$ under weak selection for $N$ large enough if
  \begin{align}\label{onethird1}
\sum_{j=1}^D \pi_j(a_j-c_j) +2\sum_{j=1}^D \pi_j(b_j-d_j) >0.
\end{align}
This extends the one-third law of evolution for a large well-mixed population (Nowak \emph{et al.}, 2004 \cite{*NSTF2004}) to a population subdivided into multiple large differentiated demes by weighing the payoffs within the demes with their reproductive values.

In a similar way, for the frequency of $A$ in Section 4 under soft selection in a large number of demes and a Wright-Fisher reproduction scheme, Eq. (\ref{appro10}) leads to 
\begin{align}\label{}
\mathbb{E}(\Delta x(\tau) \, | \, \mathbf{z}(\tau))
&= s\sum_{k=0}^N z_k(\tau) x(k)(1-x(k))((a-c)x(k) + (b-d)(1-x(k)))+o(s).
\end{align}
This is for a current population state $\mathbf{z}(\tau)=(z_0(\tau), z_1(\tau), \ldots, z_N(\tau))^T$, which gives the frequencies of demes in which the frequency of $A$ is $x(k)=k/N$ for $k=0, 1, \ldots, N$, and a corresponding frequency of $A$ in the population  $x(\tau) =\sum_{k=0}^N x(k)z_k(\tau)$. Then, the first-order effect of selection on $\rho_A$ given an initial frequency $1/(ND)$ is
 \begin{align}
 s\left( (a-c) \mathbb{E}_0\left(\sum_{k=0}^N\sum_{\tau=0}^{\infty} z_k(\tau)x(k)^2 (1-x(k))\right)+(b-d) \mathbb{E}_0\left(\sum_{k=0}^N\sum_{\tau=0}^{\infty} z_k(\tau)x(k) (1-x(k))^2\right)\right),
\end{align}
where
  \begin{align}
 \mathbb{E}_0\left(\sum_{k=0}^N\sum_{\tau=0}^{\infty} z_k(\tau)x(k) (1-x(k))^2\right)
&\approx \frac{1}{ND}f_{33} \mathbb{E}_0(S_3 ) +\frac{1}{3ND}(f_{33} + f_{32}) \mathbb{E}_0(S_2 )\nonumber\\
&\approx\frac{1}{3ND} (2f_{33} + f_{32})\mathbb{E}_0(S_2 ) 
\end{align}
and
  \begin{align}
 \mathbb{E}_0\left(\sum_{k=0}^N\sum_{\tau=0}^{\infty} z_k(\tau)x(k)^2 (1-x(k))\right)
&\approx \frac{1}{3ND}(f_{33} +f_{32}) \mathbb{E}_0(S_2 )
\end{align}
for $D$ large enough. Here, recall that $f_{32}$ and $f_{33}$ represent  the probabilities for  three offspring chosen at random in the same deme before dispersal to have two and three ancestors in different demes, respectively, in the case of an infinite number of demes in the absence of selection (see Appendix E).  We conclude $\rho_A>1/(ND)$ under weak selection for $D$ large enough if
  \begin{align}\label{onethird2}
(f_{33} +f_{32}) (a-c) + (2f_{33} +f_{32}) (b-d) >0.
\end{align}
This is the  extension of the one-third law of evolution to Wright's finite-island model (Lessard, 2011b \cite{*L2011b}).

\section{Low-migration limit}

In this section, we assume $m_{ii}=1-m+ m\alpha_{ii}$ and $m_{ij}=m\alpha_{ij}$ for $i,j =1,\ldots, D$ with $j\ne i$ in the model of Section 2. 
The state space for the frequency of $A$ in the $D$ demes of the population is represented by
\begin{equation}
E_{}=\left\{\mathbf{x}=(x_1,\ldots,x_D): x_i \in \{0, 1/N, \ldots, (N-1)/N, 1\} \textrm{ for } i=1, \ldots, D\right\},
\end{equation}
The frequency of $A$ in deme $i$ after soft selection and migration of offspring is given by
\begin{align}
\tilde{\tilde{x}}_i=\tilde{x}_i + m\sum_{j=1}^D \alpha_{ij}( \tilde{x}_j-\tilde{x}_i)
\end{align}
for $i=1, \ldots, D$. Following binomial sampling within demes according to a Wright-Fisher updating rule, the transition matrix for the population state from one time step to the next can be written in the form
 \begin{equation}\label{}
 \mathbf{S}_{}=\mathbf{U}+m\mathbf{V}(m),
 \end{equation}
 where $\mathbf{U}$ is the transition matrix in the absence of migration and 
$\lim_{m\rightarrow 0}\mathbf{V}(m) = \mathbf{V}=(v_{\mathbf{x}, \mathbf{x}' })_{\mathbf{x}, \mathbf{x}' \in E} $. 
Owing to M\"ohle's (1998a \cite{*M1998a}) lemma (see Appendix B), we have
  \begin{equation}
\lim_{m \rightarrow0}\left( \mathbf{U}+m\mathbf{V}(m)\right)^{\lfloor t/m\rfloor}=\mathbf{P}\exp\{t\mathbf{PVP}\},
\end{equation}
where $\lfloor t/m\rfloor$ denotes the floor value of $t/m$ for $t>0$ and $ \mathbf{P}=\lim_{\tau \rightarrow\infty}\mathbf{U}^{\tau}=( \rho_{\mathbf{x,} \mathbf{x}^{\prime}})_{\mathbf{x}, \mathbf{x}' \in E}$. Here, the non-null entries of $ \mathbf{P}$ are
  \begin{equation}\label{}
 \rho_{\mathbf{x,} \mathbf{x}^{\prime}}=\prod_{i=1}^D \rho_{i, x_i, x^{\prime}_i}
\end{equation}
for $ \mathbf{x}^{\prime}=(x^{\prime}_1, \ldots, x^{\prime}_D) \in \{0, 1\}^D$, where $\rho_{i, x_i, x^{\prime}_i}$ represents the probability of fixation or extinction of $A$ in deme $i$ ($x^{\prime}_i=1$ or $x^{\prime}_i=0$) starting from a frequency $x_i$ in the absence of migration, for $i=1, \ldots, D$. Then, the non-null entries of $\mathbf{G}=\mathbf{PVP}=(g_{\mathbf{x}, \mathbf{x}^{\prime}})_{\mathbf{x}, \mathbf{x}' \in E}$ are
 \begin{equation}\label{}
 g_{\mathbf{x}, \mathbf{x}^{\prime}}=\sum_{\mathbf{x}^{\prime\prime\prime}\in \{0, 1\}^D}\rho_{\mathbf{x}, \mathbf{x}^{\prime\prime\prime}}\sum_{\mathbf{x}^{\prime \prime}\in E}v_{\mathbf{x}^{\prime\prime\prime}, \mathbf{x}^{\prime \prime}}\rho_{\mathbf{x}^{\prime \prime}, \mathbf{x}^{\prime}}
 \end{equation}
 for $\mathbf{x}\in E$ and $\mathbf{x}^{\prime}\in \{0, 1\}^D$. 
 
 As $m\rightarrow 0$ with $m^{-1}$ time steps as unit of time and any initial population state $\mathbf{x}\in E$, there is an instantaneous transition to $\mathbf{x}^{\prime}\in \{0, 1\}^D$ with probability $\rho_{\mathbf{x}, \mathbf{x}^{\prime}}$. Moreover, this is followed  by a continuous-time Markov chain on $\{0, 1\}^D$ whose infinitesimal generator is  $( g_{\mathbf{x,} \mathbf{x}^{\prime}})_{\mathbf{x}, \mathbf{x}' \in \{0, 1\}^D}$. Its entries off the main diagonal are given by
  \begin{align}\label{}
 g_{\mathbf{x}, \mathbf{x}^{\prime}} &=\sum_{\mathbf{x}^{\prime \prime}\in E}v_{\mathbf{x}^{}, \mathbf{x}^{\prime \prime}}\rho_{\mathbf{x}^{\prime \prime}, \mathbf{x}^{\prime}}\nonumber\\ 
 &= N\sum_{i=1}^D(1-x_i)\left(\sum_{j=1}^D \alpha_{ij} x_j \right)\rho_{\mathbf{x}+\mathbf{e}_i/N, \mathbf{x}^{\prime}}+N\sum_{i=1}^Dx_i\left(1-\sum_{j=1}^D \alpha_{ij} x_j \right)\rho_{\mathbf{x}-\mathbf{e}_i/N, \mathbf{x}^{\prime}}
 \end{align}
 for $\mathbf{x}, \mathbf{x}' \in \{0, 1\}^D$ with $\mathbf{x}\ne  \mathbf{x}'$, where $\mathbf{e}_i$ denotes a vector with $1$ in the $i$-th coordinate and $0$ everywhere else for $i=1, \ldots, D$.
The only entries off the main diagonal with non-null values are
  \begin{align}\label{}
 g_{\mathbf{x}, \mathbf{x}^{}+\mathbf{e}_i} &=N\left(\sum_{j=1}^D \alpha_{ij} x_j \right)\rho_A(i)
 \end{align}
 if $x_i=0$, where $\rho_A(i)=\rho_{i, 1/N, 1}$, and
 \begin{align}\label{}
 g_{\mathbf{x}, \mathbf{x}^{}-\mathbf{e}_i } &=N\left(1-\sum_{j=1}^D \alpha_{ij} x_j \right)\rho_B(i)
 \end{align}
  if $x_i=1$, where $\rho_B(i)=\rho_{i, 1-1/N, 0}$. Moreover, owing to the previous section and the fact that $\mathbb{E}_0(S_2 )\approx N$ in a neutral Wright-Fisher population of size $N$, we have the approximations
  \begin{align}\label{}
\rho_A(i)\approx  \frac{1}{N} + \frac{s}{3} \left( (a_i-c_i) +2 (b_i-d_i)\right)
 \end{align}
 \begin{align}\label{}
\rho_B(i) \approx \frac{1}{N} + \frac{s}{3} \left( (d_i-b_i) +2 (c_i-a_i)\right)
 \end{align}
 for selection weak enough and $N$ large enough, for $i=1, \ldots, D$.
 
 In the case of uniform dispersal, that is, $\alpha_{ij}=1/D$ for $i,j=1, \ldots, D$, the frequency of $A$ in the whole population is given by $x=\sum_{j=1}^D \alpha_{ij} x_j$. Moreover, it is a birth-death process on the state space $\{0, 1/D, \ldots, (D-1)/D, 1\}$ with birth rate $Dx(1-x)\sum_{i=1}^D \rho_A(i)$ and death rate $Dx(1-x)\sum_{i=1}^D \rho_B(i)$. Like in the Gambler's ruin problem (see, e.g., Karlin and Taylor, 1975 \cite{*KT1975}), there is fixation of $A$ in the whole population with probability
   \begin{align}\label{}
\rho_A = \frac{1-\frac{\sum_{i=1}^D \rho_B(i)}{\sum_{i=1}^D \rho_A(i)}}{1-\left(\frac{\sum_{i=1}^D \rho_B(i)}{\sum_{i=1}^D \rho_B(i)}\right)^D} \approx \frac{1}{D}+ \frac{N(D-1)}{2D} \left( \bar{a}-\bar{c} +\bar{b}- \bar{d}\right)s,
 \end{align}
 where
    \begin{align}\label{}
\bar{a} = \frac{1}{D}\sum_{i=1}^D a_i, \quad \bar{b} = \frac{1}{D}\sum_{i=1}^D b_i, \quad \bar{c} = \frac{1}{D}\sum_{i=1}^D c_i, \quad \bar{d} = \frac{1}{D}\sum_{i=1}^D d_i.
 \end{align}
 We conclude that $\rho_A > 1/D$ under weak selection if 
    \begin{align}\label{}
 (\bar{a}-\bar{c}) + (\bar{b}- \bar{d})>0.
 \end{align}
 This means that $A$ is risk-dominant over $B$ with respect to the average payoffs in pairwise interactions in the whole population.

\section{Low-mutation limit}

In this section, we assume that there is mutation from one type to the other type with probability $u >0$ for every offspring after selection and migration independently of all the others in the model of Section 2 under soft selection and reproduction of the Wright-Fisher type. 

For the population state in the space $E$ defined in the previous section,
the transition matrix from one time step to the next can be decomposed into
 \begin{equation}\label{}
 \mathbf{T}_{}=\mathbf{S}+u\mathbf{R}(u),
 \end{equation}
 where $\mathbf{S}$ is the transition matrix in the absence of mutation and $\mathbf{R}(u) \rightarrow \mathbf{R}=(r_{\mathbf{x}, \mathbf{x}^{\prime}})_{\mathbf{x}, \mathbf{x}' \in E}$ as $u \rightarrow 0$. Only the entries in the first and last rows of the matrix $\mathbf{R}$ corresponding to the fixation states $\mathbf{0}$ and $\mathbf{1}$ will matter in the limit below.
 The first row is given by
  \begin{equation}\label{}
r_{\mathbf{0}, \mathbf{0}}=-DN, \quad \textrm{} r_{\mathbf{0}, \mathbf{e}_i/N}=N \quad \textrm{for } i=1, \ldots, D,
 \end{equation}
 and the last one by
 \begin{equation}\label{}
r_{\mathbf{1}, \mathbf{1}}=-DN, \quad \textrm{} r_{\mathbf{1}, \mathbf{1}-\mathbf{e}_i/N}=N \quad \textrm{for } i=1, \ldots, D.
 \end{equation}
 Owing to the above decomposition, M\"ohle's (1998a \cite{*M1998a}) lemma  (see Appendix B) guarantees that
 \begin{equation}
\lim_{u \rightarrow0}\left( \mathbf{S}+u\mathbf{R}(u)\right)^{\lfloor t/u\rfloor}=\mathbf{P}\exp\{t\mathbf{PRP}\}
\end{equation}
for $t>0$, where the non-null entries of $ \mathbf{P}=\lim_{\tau\rightarrow \infty} \mathbf{S}^{\tau}=( \rho_{\mathbf{x,} \mathbf{x}^{\prime}})_{\mathbf{x}, \mathbf{x}' \in E}$ are $\rho_{\mathbf{x}, \mathbf{0}}$ and $\rho_{\mathbf{x}, \mathbf{1}}$.
These are the fixation probabilities of $\mathbf{0}$ and $\mathbf{1}$, respectively, given an initial population state $\mathbf{x}$. Moreover, the non-null entries of $\mathbf{G}=\mathbf{PRP}=(g_{\mathbf{x}, \mathbf{x}^{\prime}})_{\mathbf{x}, \mathbf{x}' \in E}$ are
 \begin{equation}\label{}
 g_{\mathbf{x}, \mathbf{0}^{}}=\rho_{\mathbf{x}, \mathbf{0}}\sum_{\mathbf{x}^{\prime}\in E}r_{\mathbf{0}, \mathbf{x}^{\prime}}\rho_{\mathbf{x}^{\prime}, \mathbf{0}^{}}+\rho_{\mathbf{x}, \mathbf{1}}\sum_{\mathbf{x}^{\prime}\in E}r_{\mathbf{1}, \mathbf{x}^{\prime}}\rho_{\mathbf{x}^{\prime}, \mathbf{0}^{}}
\end{equation}
and
 \begin{equation}\label{}
 g_{\mathbf{x}, \mathbf{1}^{}}=\rho_{\mathbf{x}, \mathbf{0}}\sum_{\mathbf{x}^{\prime}\in E}r_{\mathbf{0}, \mathbf{x}^{\prime}}\rho_{\mathbf{x}^{\prime}, \mathbf{1}^{}}+\rho_{\mathbf{x}, \mathbf{1}}\sum_{\mathbf{x}^{\prime}\in E}r_{\mathbf{1}, \mathbf{x}^{\prime}}\rho_{\mathbf{x}^{\prime}, \mathbf{1}^{}}.
\end{equation}
In the limit of a low mutation probability $u$ with $u^{-1}$ time steps as unit of time and any initial population state $\mathbf{x}$, there is an instantaneous transition to $\mathbf{0}$ with probability $\rho_{\mathbf{x}, \mathbf{0}}$ or to $\mathbf{1}$ with probability $\rho_{\mathbf{x}, \mathbf{1}}$. Moreover, this is followed  by a continuous-time Markov chain on these two states whose infinitesimal generator is  
 \begin{equation}\label{}
\begin{pmatrix}
g_{\mathbf{0}, \mathbf{0}^{}}&g_{\mathbf{0}, \mathbf{1}^{}} \\
g_{\mathbf{1}, \mathbf{0}^{}}&g_{\mathbf{1}, \mathbf{1}^{}}  \\
\end{pmatrix},
\end{equation}
where
\begin{equation}\label{}
 g_{\mathbf{0}, \mathbf{1}^{}}=N \sum_{i=1}^D\rho_{\mathbf{e}_i/N, \mathbf{1}}=-g_{\mathbf{0}, \mathbf{0}^{}}
\end{equation}
and
\begin{equation}\label{}
 g_{\mathbf{1}, \mathbf{0}^{}}=N \sum_{i=1}^D\rho_{\mathbf{1-}\mathbf{e}_i/N, \mathbf{0}}=-g_{\mathbf{1}, \mathbf{1}^{}}.
\end{equation}
Owing to the theory of Markov chains (see, e.g., Karlin and Taylor, 1975 \cite{*KT1975}), the time-average abundance of strategy $A$ in the long run is given by
\begin{equation}\label{}
 \frac{g_{\mathbf{0}, \mathbf{1}^{}}}{g_{\mathbf{0}, \mathbf{1}^{}}+g_{\mathbf{1}, \mathbf{0}^{}}}=\frac{\rho_A}{\rho_A+\rho_B},
\end{equation}
where
\begin{equation}\label{}
 \rho_A=\frac{1}{D}\sum_{i=1}^D\rho_{\mathbf{e}_i/N, \mathbf{1}}
\end{equation}
and
\begin{equation}\label{}
 \rho_B=\frac{1}{D}\sum_{i=1}^D\rho_{\mathbf{1-}\mathbf{e}_i/N, \mathbf{0}}
\end{equation}
are  the fixation probabilities of $A$  and $B$, respectively,  introduced as single mutants.
This result extends a result proved and applied in Fudenberg and Imhof (2006 \cite{*FI2006}) and Fudenberg \emph{et al.}, 2006 \cite{*FNTI2006}) to the case of multiple demes.

\section{Discussion}

The existence of two timescales has been applied for some time in evolutionary game theory but mathematical arguments to justify their application are often lacking. In this paper, we have used two convergence results for discrete-time Markov chains, namely, Ethier and Nagylaki's (1980 \cite{*EN1980}) theorem and Mohle's (1998a \cite{*M1998a}) lemma, to deduce strong- and low-migration limits of linear games, besides a low-mutation limit, in a population subdivided into a finite number of finite demes. 

With population-scaled unit of time and payoffs  as the deme size tends to infinity, the strategy frequencies in the whole population are described by a diffusion process whose infinitesimal mean and variance depend on reproductive values of the demes in the absence of selection. 
Actually, an effective game matrix in a well-mixed population that is an average of the payoff matrices within demes with respect to these reproductive values leads to the same dynamics (see Eq. (\ref{matrix1})). 

With the same payoff matrix within demes and partial uniform dispersal as the number of demes tends to infinity, the effective game matrix involves probabilities for up to three offspring within demes to have one, two or three ancestors in a stationary population of an infinite number of demes in the absence of selection. This is in agreement with results on the deterministic dynamics (Ohtsuki, 2010 \cite{*O2010}, Lessard, 2011a \cite{*L2011a}). In Eq. (\ref{matrix2}), the probabilities $f_{21}$ and $f_{31}$ correspond to identity-by-descent (IBD) measures for two individuals in interaction, called actor and recipient to distinguish them, 
and for these two individuals and a third one in competition with the actor. As for the probability $f_{32}/3$, it stands for the situation where either only the competitor and the actor are IBD, or only the competitor and the recipient are IBD. The positive terms in Eq. (\ref{matrix2}) represent effects of interaction and the negative terms effects of competition. Global (hard) selection instead of local (soft) selection introduces extra terms in the effective matrix (see Eq. (\ref{matrix3})). These diminish the effects of competition by a factor $(1-m)^2$, where $m$ is the dispersal probability. The diminution is negligible when $m$ is small but eliminates competition completely  when $m=1$. Simulations and some analytical results suggest that a diffusion approximation holds in
the many-demes limit under a variety of assumptions such as isolation by distance, local extinction and diploid population (Whitlock 2003 \cite{*W2003}, Roze and Rousset 2003 \cite{*RR2003}, Lessard, 2009 \cite{*L2009}).

Several evolutionary properties of strategies in finite populations are based on the fixation probability of a single mutant (see, e.g., Rousset and Billiard, 2000 \cite{*RB2000}, Rousset, 2003 \cite{*R2003}, Nowak \emph{et al.,} 2004 \cite{*NSTF2004}). A diffusion approximation for the strategy frequencies in the limit of a large population allows us to use a well-known formula  that depends only on the infinitesimal mean and variance (see, e.g., Ewens, 2004 \cite{*E2004}). Then, the first-order effect of selection on the fixation probability can be calculated by letting the population-scaled intensity of selection go to zero. Alternatively, this effect can be obtained directly by considering successive expected changes in strategy frequencies that involve expected coalescence times of ancestral lines under neutrality (Rousset, 2003 \cite{*R2003}, Lessard and Ladret, 2007 \cite{*LL2007}). These can be approached in a subdivided population as the size or number of demes tends to infinity by resorting to the existence of two timescales in the genealogical process under neutrality with a population-scaled unit of time. At least in the case of updating according to a Wright-Fisher binomial scheme or a Moran type model that leads to a diffusion approximation, this process tends to the Kingman coalescent (Kingman, 1982 \cite{*K1982}) with pairs of ancestral lines coalescing at rate $1$ independently of one another after instantaneous initial transitions. This might be the case for a wide range of population structures (Allen and McAvoy, 2024 \cite{*AM2024}).

The conditions for the fixation probability of a strategy $A$ introduced as a single mutant to exceed its initial frequency in the strong-migration limits as the size or number of demes tends to infinity are extensions of the one-third law of evolution (Nowak \emph{et al.}, 2004 \cite{*NSTF2004}). This law for a linear game in a well-mixed population states that this is the case if the average payoff to the mutant strategy is the largest one when its frequency is equal to $1/3$. In the strong-migration limit under soft selection as the deme size tends to infinity, this law holds for weighted strategy frequencies and weighted average payoffs with weights given by the reproductive values of the demes under neutrality (see Eq. (\ref{onethird1})). In the strong-migration limit under soft selection as the number of demes tends to infinity with the same payoff matrix within demes of the same fixed size $N$ and uniform dispersal of offspring with probability $m$, the ratio $1/3$ is replaced by 
\begin{equation}\label{}
\frac{f_{33} +f_{32} }{3f_{33} +2f_{32}}\approx \frac{2\nu^2+3\nu}{6\nu^2+6\nu}> \frac{1}{3},
  \end{equation}
  where $\nu=Nm$. Actually, the ratio decreases from $1/2$ to $1/3$ as $\nu$ increases from $0$ to infinity. The condition with a ratio $1/2$ corresponds to risk dominance. This condition is less stringent than the condition with a ratio $1/3$ when we have the inequalities $a>c>d>b$ for the payoffs to $A$ and $B$ against $A$ and $B$, respectively. This is the case, for instance, in a stag-hunt game or a Prisoner's dilemma  repeated enough time with tit-for-tat and always-defect as strategies (see, e.g., Nowak \emph{et al.}, 2004 \cite{*NSTF2004}, Archetti and Scheuring, 2012 \cite{*AS2012}).
 
 In the low-migration limit with the inverse of the intensity of migration as unit of time, there is initially instantaneous fixation within demes of size $N$ and, after that, each deme receives single migrants from the other demes at given rates. Moreover, if deme $i$ was fixed for $B$, then a single migrant of type $A$ instantaneously fixes within the deme with probability $\rho_A(i)$, for $i=1, \ldots, D$, and analogously for a migrant of type $B$ in a deme fixed for $A$. In the particular case of uniform dispersal,  the fixation probability of $A$ in the whole population after being introduced as a single mutant exceeds its initial frequency under the effect of weak selection when $A$ is risk dominant over $B$ with respect to the average payoff matrix in the whole population. Although the one-third law of evolution holds locally within demes, it does not hold globally in the entire population. This is an important limitation since some form of viscosity affecting migration is expected to exist in spatially structured populations. Note that a diffusion approximation can be obtained in the low-migration limit as the deme size goes to infinity (Slatkin, 1981 \cite{*S1981}).

Finally, we have shown that the fixation probabilities of $A$ and $B$ introduced as single mutants in a deme chosen at random determine their average abundance in the long run under weak recurrent mutation. This has previously been used (see, e.g., Rousset and Billiard, 2000 \cite{*RB2000}) and proved (Fudenberg and Imhof, 2006 \cite{*FI2006}) in simpler contexts, but the use of M\"ohle's (1998a \cite{*M1998a}) lemma simplifies the proof and makes extensions possible. This is also the case for the low-migration limit which can be studied directly for a Moran type model (Pires and Broom, 2024 \cite{*PB2024}), but becomes more complicated under other assumptions such as a Wright-Fisher updating rule.

As for diffusion approximations of discrete-time Markov chains with two timescales, we have apply Ethier and Nagylaki's (1980 \cite{*EN1980}) theorem to ascertain the strong-migration limits as the size or number of demes goes to infinity. However, the conditions to check  are rather technical and we had to resort to previous papers for the most technical ones (Nagylaki, 1980 \cite{*N1980}, Wakeley, 2003 \cite{*W2003}, Lessard, 2007 \cite{*L2007}, Lessard, 2009 \cite{*L2009}, Kroumi and Lessard, 2024a \cite{*KL2024a}). The key condition that the deterministic recurrence system in an infinite population in the absence of selection 
converges globally is connected to the ergodic theorem for Markov chains. However, it goes beyond it in the strong-migration limit as the number of demes tends to infinity. In this case, the condition relies on coalescent theory (Kingman, 1982 \cite{*K1982}; see Wakeley, 2009 \cite{*W2009}) and is related to an extension of the Ewens sampling formula (Ewens, 1972 \cite{*E1972}; see Ewens, 2004 \cite{*E2004}) for a finite population with mutation to a new allele replaced by migration to a new deme (Wakeley, 2003 \cite{*W2003}, Lessard, 2007 \cite{*L2007}, 2009 \cite{*L2009}). The same approach can be used under various assumptions such as random payoffs and extinction of demes  followed by recolonization (Kroumi and Lessard, 2024b \cite{*KL2024b}). The case where the extinction of a deme would depend on its composition is more complicated. With demes of size $2$ and only two strategies, the recurrence system under neutrality is one-dimensional and global convergence can be shown as in models for the evolution of cooperation with an opting-out strategy (Zhang \emph{et al.}, 2016 \cite{*zha82016}, Li and Lessard, 2021 \cite{*LL2021}). With multiple strategies, not to mention demes of size greater than $2$, the analysis is challenging. Global convergence in continuous-time models (Cressman and K\v{r}ivan, 2022 \cite{*CK2022}) suggests that this could not be out of reach. Evolutionary games in class-structured populations such as age-structured populations (see, e.g., Lessard and Soares, 2018 \cite{*LS2018}, Soares and Lessard 2020 \cite{*SL2020}, Priklopil and Lehman 2024 \cite{*PL2024}) raise similar challenges.

A diffusion approximation in the limit of a large population when the payoffs are proportional to the inverse of the population size yields a useful formula for the fixation probability of a single mutant strategy that can be calculated for small population-scaled payoffs. This is not necessary, however, to get the first-order effect of selection on the fixation probability. This effect can be obtained directly by considering the first-order effect of selection on the sum of the successive expected changes in the mutant type frequency and by using  coalescent theory in the absence of selection. The two timescales in the genealogical process can be handled  owing to M\"ohle's (1998a \cite{*M1998a}) lemma. This approach can deal easily with complications such as genetic recombination (Lessard and Wakeley, 2004 \cite{*LW2004}), multiple types (Antal \emph{et al.}, 2009 \cite{*ANT2009}, Lessard and Lahaie, 2009 \cite{*LL2009}) or non-linear average payoffs (Lessard and Ladret, 2007 \cite{*LL2007}). It is worth noting that the genealogical process comes also into play in the diffusion approximation since the infinitesimal mean involves identity measures between interacting and competing individuals. Moreover, it can even deal with models that are out of reach of a diffusion approximation because of assumptions such as a highly skewed distribution in offspring production which can have a big impact in genetic variability (Lessard and Ladret, 2007 \cite{*LL2007}, Lessard, 2011b \cite{*L2011b}, Wilton \emph{et al.}, 2017 \cite{*W2017}, Diamantidis \emph{et al.}, 2024 \cite{*DW2024}).

Finally, M\"ohle's (1998a \cite{*M1998a}) lemma can be applied forward in time for a population with a constant finite size as some parameters such as migration or mutation probabilities tend to zero. A wide range of assumptions about the population structure besides those considered in this paper could be considered.

\section*{Appendix A: Diffusion approximation with two timescales}

Consider a finite population subdivided into demes of $L$ types with $x_l$ being the frequency of individuals of type $A$ in a deme of type $l$ for $l=1, \ldots, L$. We let $x=\sum_{l=1}^L z_l x_l$ be the frequency of $A$ in the whole population in state $\mathbf{z}=(z_1, \ldots, z_L)$. Time is discrete and changes in population states are  Markovian. Moreover, from one time step to the next, we suppose that there exist $\Delta t >0$ and an approximative population state $\mathbf{z}(x)$ such that the changes in $x$ and $\mathbf{y}=\mathbf{z}-\mathbf{z}(x)$, represented by $\Delta x$ and $\Delta \mathbf{y}$, have conditional moments that verify
\begin{subequations}\label{}
\begin{align}\label{}
&\mathbb{E}(\Delta x \, | \, \mathbf{z})=b(x, \mathbf{y})\Delta t + o(\Delta t), \\
&\mathbb{V}(\Delta x \, | \, \mathbf{z})=a(x, \mathbf{y})\Delta t + o(\Delta t), \\
&\mathbb{E}((\Delta x)^4 \, | \, \mathbf{z})=o(\Delta t), \\
&\mathbb{E}(\Delta \mathbf{y }\, | \, \mathbf{z})=c(x, \mathbf{y})+ o(1), \\
&\mathbb{V}(\Delta \mathbf{y} \, | \, \mathbf{z})=o(1),
\end{align}
\end{subequations}
with 
\begin{align}\label{}
 \mathbf{Y }^{(k+1)}=  \mathbf{Y}^{(k)}+c(x, \mathbf{Y}^{(k)})\rightarrow \mathbf{0}
\end{align}
as $k \rightarrow \infty$, for some functions $a(x, \mathbf{y})$, $b(x, \mathbf{y}) $ and $c(x, \mathbf{y})$. Then, under usual regularity and uniformity conditions, and with $(\Delta t)^{-1}$ time steps as unit of time, the frequency of $A$ in the population as $\Delta t \rightarrow 0$ converges weakly to a continuous-time diffusion process with $m(x)= b(x, \mathbf{0})$ as infinitesimal mean and $v(x)= a(x, \mathbf{0})$ as infinitesimal variance. 

This is a special version of the convergence result for Markov chains with two timescales that can be found in Ethier and Nagylaki (1980 \cite{*EN1980}) by setting $m=1$ and $\Delta_N=1$ in there. 

%

\section*{Appendix B: Transition matrix with two timescales}

Suppose a transition matrix for a finite-time Markov chain on a finite state space that can be expressed into the form
 \begin{equation}
\mathbf{Q}=\mathbf{A}+c_n\mathbf{B}_n,
\end{equation}
where $\mathbf{A}$ is a transition matrix such that $\lim_{\tau\rightarrow \infty}\mathbf{A}^{\tau}=\mathbf{P}$, while $\lim_{n\rightarrow \infty}\mathbf{B}_n=\mathbf{B}$ and $\lim_{n\rightarrow \infty}c_n=0$ with $c_n>0$ for $n\geq 1$. Then, we have
 \begin{equation}
\lim_{n\rightarrow\infty}\left( \mathbf{A}+c_n\mathbf{B}_n\right)^{\lfloor t/c_n\rfloor}=\mathbf{P}\exp\{t\mathbf{PBP}\},
\end{equation}
where $\lfloor t/c_n\rfloor$ denotes the floor value of $t/c_n$ for $t>0$. The matrix $\mathbf{PBP}$ is the infinitesimal generator of a continuous-time Markov chain that takes place after  initial instantaneous transitions according to probabilities given by the entries of the projection matrix $\mathbf{P}$. See M\"ohle (1998a \cite{*M1998a}) for a proof.

\section*{Appendix C: Genealogical process in the neutral Wright-Fisher population structured into large demes}

In this appendix, we derive the backward-time genealogical process of a sample taken from a population structured into $D$ demes under the neutral Wright-Fisher model in the limit of a large deme size $N$ which was considered by Notohara (1990 \cite{*N1990}, 1993 \cite{*N1993}).

We consider a sample of size $n\ge2$ at a given time step. Looking backward in time at the genealogy of this sample, the distribution of the ancestors in the $D$ demes at any previous time step can be described by a vector
\begin{equation}
\mathbf{n}=(n_1,\ldots,n_D),
\end{equation}
where $n_i$ denotes the number of ancestors in deme $i$, for $i=1,\ldots,D$. Then
\begin{equation}|\mathbf{n}|=n_1+\cdots+n_D
\end{equation}
is the total number of ancestors. 

Let $\mathbf{n}(\tau)$ be the distribution of the ancestors in the $D$ demes $\tau \geq 0$ time steps back. Given an initial sample of $n$ individuals, this is a discrete-time Markov chain with state space
\begin{equation}
S_{}=\left\{\mathbf{n}=(n_1,\ldots,n_D)\in \mathbb{N}^D : 1\le n_1+\cdots+n_D\le n\right\}.
\end{equation}
This sample set can be decomposed into sample subsets according to the  number of ancestors, that is,
\begin{equation}
S_{}=\cup_{k=1}^n S_k,
\end{equation}
where
\begin{equation}S_{k}=\{\mathbf{n}=(n_1,\ldots,n_D)\in \mathbb{N}^D : |\mathbf{n}|=n_1+\cdots+n_D=k\}
\end{equation}
for $k=1,\ldots,n$. We assume throughout this section that the states are ordered so that those in $S_1$ come first, those in $S_2$ come next, and so on up to those in $S_n$. 

Now, let $\boldsymbol{Q}=\left(Q_{\mathbf{n},\mathbf{n}'}\right)_{\mathbf{n},\mathbf{n}'\in S}$ be the transition matrix of the above genealogical process from one time step to the previous one. This stochastic matrix can be written in the form
 \begin{equation}\label{}
 \mathbf{Q}_{}=\mathbf{A}+\frac{\mathbf{B}}{N}+ \frac{\mathbf{O}(1)}{N^2}, 
 \end{equation}
where the entries of $\mathbf{O}(1)$ are bounded as $N \rightarrow \infty$, while $\mathbf{A}$ and $\mathbf{B}$ are in the block forms
\begin{equation}
\mathbf{A}= \begin{pmatrix}
\mathbf{A}_1&\mathbf{0}&\ldots&\mathbf{0} \\
\mathbf{0}&\mathbf{A}_2&\ldots&\mathbf{0} \\
\vdots&\vdots&\ddots&\vdots\\
\mathbf{0}&\mathbf{0}&\ldots&\mathbf{A}_n \\
\end{pmatrix}
\end{equation}
and
\begin{equation}\label{}
\mathbf{B}= \begin{pmatrix}
\mathbf{0}&\mathbf{0}&\mathbf{0}&\ldots&\mathbf{0} &\mathbf{0} \\
\mathbf{B}_{2,1}&\mathbf{B}_{2,2}&\mathbf{0}&\ldots&\mathbf{0}&\mathbf{0}  \\
\mathbf{0}&\mathbf{B}_{3,2}&\mathbf{B}_{3,3}&\ldots&\mathbf{0}&\mathbf{0}  \\
\vdots&\vdots&\vdots&\ddots&\vdots&\vdots  \\
\mathbf{0}&\mathbf{0}&\mathbf{0}&\ldots&\mathbf{B}_{n,n-1}&\mathbf{B}_{n,n}  \\
\end{pmatrix}
\end{equation}
with respect to the sample subsets $S_1, \ldots, S_n$.
Here,
$\mathbf{0}$ denotes a matrix of any dimension whose all entries are zero. The entries of $\mathbf{A}_k=(a_{\mathbf{n}, \mathbf{n}'})_{\mathbf{n},\mathbf{n}'\in S_k}$ are transition probabilities from states in $S_k$ to states in $S_k$ for $k=1, \ldots, n$ as if the demes would be of infinite size so that the number of ancestors remains constant. In this case, the genealogical process is described by independent Markov chains whose transition matrix is $\mathbf{M}$ and stationary distribution $\boldsymbol{\pi}^T=(\pi_1, \ldots, \pi_D)$. Then, in the case of $k$ ancestors, the stationary distribution is given by
 \begin{equation}\label{}
\pi_{\mathbf{n}}=\frac{|\mathbf{n}|!}{n_1!\times n_2!\times\cdots\times n_d!}\pi_{1}^{n_1}\times\cdots\times \pi_{d}^{n_d}
 \end{equation}
 for $\mathbf{n} \in S_k$ and, owing to the ergodic theorem (see, e.g., Karlin and Taylor 1975 \cite{*KT1975}), we have 
 \begin{align}
\lim_{\tau \rightarrow \infty}\mathbf{A}_k^{\tau} = \mathbf{P}_k =\mathbf{1} (\pi_{\mathbf{n}})_{{\mathbf{n}\in S_k}}
\end{align}
 for $k=1, \ldots, n$.
As for the non-null entries of $\mathbf{B}_{}=(b_{\mathbf{n}, \mathbf{n}'})_{\mathbf{n}, \mathbf{n}'\in S_{}}$, they are given by
\begin{equation}\label{}
b_{\mathbf{n}, \mathbf{n}'}=-\sum_{i=1}^D \frac{(n_i^\prime-1) n_i^\prime}{2} a_{\mathbf{n}, \mathbf{n}'}
 \end{equation}
 if $\mathbf{n}, \mathbf{n}' \in S_k$, and
 \begin{equation}\label{}
b_{\mathbf{n}, \mathbf{n}'}=\sum_{i=1}^D \frac{n_i^\prime (n_i^\prime +1)}{2} a_{\mathbf{n}, \mathbf{n}'+\boldsymbol{\mathrm{e}}_i}
 \end{equation}
 if $\mathbf{n}\in S_k$ and $\mathbf{n}' \in S_{k-1}$, where $\boldsymbol{\mathrm{e}}_i$ denotes a $D$-dimensional vector with $1$ in the $i$-th entry and $0$ everywhere else.
 
Under the above conditions,  M\"ohle (1998a \cite{*M1998a}) lemma (see Appendix B) guarantees that
 \begin{equation}
\lim_{N\rightarrow\infty}\left( \mathbf{A}+\frac{\mathbf{B}}{N}+ \frac{\mathbf{O}(1)}{N^2}\right)^{\lfloor tN\rfloor}=\mathbf{P}\exp\{t\mathbf{G}\},
\end{equation}
where
 \begin{equation}\label{}
  \mathbf{P}=\lim_{\tau \rightarrow\infty}\mathbf{A}^{\tau}= \begin{pmatrix}
\mathbf{P}_1&\mathbf{0}&\ldots&\mathbf{0} \\
\mathbf{0}&\mathbf{P}_2&\ldots&\mathbf{0} \\
\vdots&\vdots&\ddots&\vdots\\
\mathbf{0}&\mathbf{0}&\ldots&\mathbf{P}_n \\
\end{pmatrix}
\end{equation}
and
 \begin{equation}\label{}
\mathbf{G}=\mathbf{P}\mathbf{B}\mathbf{P}=\begin{pmatrix}
\mathbf{0}&\mathbf{0}&\mathbf{0}&\ldots&\mathbf{0} &\mathbf{0} \\
\mathbf{P}_2\mathbf{B}_{2,1}\mathbf{P}_1&\mathbf{P}_2\mathbf{B}_{2,2}\mathbf{P}_2&\mathbf{0}&\ldots&\mathbf{0}&\mathbf{0}  \\
\mathbf{0}&\mathbf{P}_3\mathbf{B}_{3,2}\mathbf{P}_2&\mathbf{P}_3\mathbf{B}_{3,3}\mathbf{P}_3&\ldots&\mathbf{0}&\mathbf{0}  \\
\vdots&\vdots&\vdots&\ddots&\vdots&\vdots  \\
\mathbf{0}&\mathbf{0}&\mathbf{0}&\ldots&\mathbf{P}_n\mathbf{B}_{n,n-1}\mathbf{P}_{n-1}&\mathbf{P}_n\mathbf{B}_{n,n}\mathbf{P}_n  \\
\end{pmatrix}.
\end{equation}
Here, we have
 \begin{equation}\label{}
\mathbf{P}_k\mathbf{B}_{k,k}\mathbf{P}_k=c_k \mathbf{1} (\pi_{\mathbf{n}})_{{\mathbf{n}\in S_k}},
\end{equation}
where
 \begin{equation}\label{}
-c_k = \sum_{\mathbf{n}' \in S_k}\sum_{i=1}^D \frac{(n_i^\prime-1) n_i^\prime}{2}\sum_{\mathbf{n}\in S_k} \pi_{\mathbf{n}}a_{\mathbf{n}, \mathbf{n}'}=\sum_{\mathbf{n}' \in S_k}\sum_{i=1}^D \frac{(n_i^\prime-1) n_i^\prime}{2}\pi_{\mathbf{n}'}=\frac{(k-1)k}{2}\sum_{i=1}^D \pi_i^2
\end{equation}
represents the expected number of pairs of ancestors  in the same deme among $k$ ancestors in the stationary state, for $k=2, \ldots, n$. This can be expressed as
 \begin{equation}\label{}
-c_k = \frac{(k-1)k}{2}\lambda
\end{equation}
for $k=2, \ldots, n$, where
 \begin{equation}\label{}
\lambda=\sum_{i=1}^D \pi_i^2.
\end{equation}
Similarly, we have
 \begin{equation}\label{}
\mathbf{P}_k\mathbf{B}_{k,k-1}\mathbf{P}_{k-1}=d_k \mathbf{1} (\pi_{\mathbf{n}})_{{\mathbf{n}\in S_{k-1}}},
\end{equation}
where
 \begin{equation}\label{}
d_k = \sum_{\mathbf{n}' \in S_{k-1}}\sum_{i=1}^D \frac{n_i^\prime(n_i^\prime+1)}{2}\sum_{\mathbf{n}\in S_k} \pi_{\mathbf{n}}a_{\mathbf{n}, \mathbf{n}'+\boldsymbol{\mathrm{e}}_i}=\sum_{\mathbf{n}' \in S_{k-1}}\sum_{i=1}^D \frac{n_i^\prime(n_i^\prime+1)}{2}\pi_{\mathbf{n}'+\boldsymbol{\mathrm{e}}_i}=\frac{(k-1)k}{2}\lambda,
\end{equation}
for $k=2, \ldots, n$.

The above analysis implies that, with
$N/\lambda$ time steps as unit of time  and after instantaneous initial transitions in the limit of a large deme size $N$, the number of ancestral lines of a sample of size $n$ decreases from $k$ to $k-1$ at the rate $(k-1)k/2$ for $k=2, \ldots, n$. This is the death process of the Kingman coalescent (Kingman, 1982 \cite{*K1982}) with every pair of ancestral lines coalescing at the rate $1$ independently of all the others.
 
\section*{Appendix D: Genealogical process in the neutral Moran population structured into large demes}

In the neutral Moran model, only one individual chosen at random in a deme chosen at random is replaced by an offspring chosen at random after migration.

 As in Kroumi and Lessard (2015 \cite{*KL2015}), the transition matrix of the genealogical process for a sample of size $n$ can be decomposed into the form
 \begin{equation}\label{}
 \mathbf{Q}_{}=\mathbf{I}+\frac{\mathbf{A}}{DN}+\frac{\mathbf{B}}{DN^2},
 \end{equation}
 where $\mathbf{I}$ is an identity matrix of size $n$, while the non-null entries of $\mathbf{A}=(a_{\mathbf{n},\mathbf{n}'})_{\mathbf{n},\mathbf{n}'\in S}$ are given by
 \begin{equation}\label{}
a_{\mathbf{n},\mathbf{n}'}=
\begin{cases}
n_im_{ji}& \text{if}\;\;\mathbf{n}'=\mathbf{n}-\mathbf{e}_{i}+\mathbf{e}_{j} \;\text{for}\;i\not=j\;\text{and}\;n_i\ge1, \\
-\sum_{i\not=j:n_i\ge1}n_im_{ji}& \text{if}\;\mathbf{n}'=\mathbf{n},
\end{cases}
\end{equation}
 and the non-null entries of $\mathbf{B}=(b_{\mathbf{n},\mathbf{n}'})_{\mathbf{n},\mathbf{n}'\in S}$ given by
\begin{equation}\label{}
 b_{\mathbf{n},\mathbf{n}'}=
\begin{cases}
n_i(n_i-1)m_{ii}+\sum_{j	\not=i}n_in_jm_{ji}& \text{if}\;\;\mathbf{n}'=\mathbf{n}-\mathbf{e}_{i}\;\text{for}\;i\;\text{such that}\;n_i\ge1, \\
-n_in_jm_{ji}& \text{if}\;\;\mathbf{n}'=\mathbf{n}-\mathbf{e}_{i}+\mathbf{e}_{j} \;\text{for}\;i\not=j\;\text{and}\;n_i\ge1, \\
-\sum_{i:n_i\ge1}n_i(n_i-1)m_{ii}& \text{if}\;\mathbf{n}'=\mathbf{n}.
\end{cases}
\end{equation}
In this case, using M\"ohle's lemma (see Kroumi and Lessard, 2015 \cite{*KL2015}, for a proof), it can be shown that
 \begin{equation}
\lim_{N\rightarrow\infty}\left( \mathbf{I}+\frac{\mathbf{A}}{DN}+\frac{\mathbf{B}}{DN^2}\right)^{\lfloor tDN^2/(2\lambda)\rfloor}=\mathbf{P}\exp\{t\mathbf{G}\},
\end{equation}
where $\mathbf{P}$ and $\mathbf{G}$ are defined as in the previous section, and 
 \begin{equation}\label{}
\lambda=\sum_{i=1}^D m_{ii}\pi_i^2+ \sum_{i=1}^D \sum_{j\ne i} m_{ij}\pi_i \pi_j=\sum_{i=1}^D m_{ii}\pi_i^2+ \sum_{i=1}^D \pi_i(\pi_i - m_{ii}\pi_i) =\sum_{i=1}^D \pi_i^2.
\end{equation}
Therefore, the same conclusion as in the Appendix C holds with $(DN)^2/(2\lambda)$ time steps as unit of time.

\section*{Appendix E: Genealogical process in the neutral Wright-Fisher population structured into many demes}

In this appendix, we will deduce a particular case of what is known as the structured coalescent (Herbots 1997 \cite{*H1997}, Nordborg, 2001  \cite{*N2001}, Wakeley, 2003 \cite{*W2003}). 

Returning to the neutral Wright-Fisher model but considering a large number of demes of fixed size $N$ with partial uniform dispersal of offspring, that is, $m_{ii}=1-m$ and $m_{ij}=m/(D-1)$ for $i,j =1,\ldots, D$ with $j\ne i$, we will represent the state $\mathbf{n}=(n_1,\ldots,n_D)$ for any given number of ancestors in the $D$ demes by a standard state
\begin{equation}
\tilde{\mathbf{n}}=(\tilde{n}_1, \ldots, \tilde{n}_d)=(n_{i_1},\ldots,n_{i_d}),
\end{equation}
where $n_{i_1}\geq \dots \geq n_{i_d}\geq 1$ and $n_{i_1}+ \dots +n_{i_d}= |\mathbf{n}|= |\tilde{\mathbf{n}}|$. The standard states are ordered so that those with at most $1$ ancestor in each deme come first (subspace $\tilde{S}_1$), those with 
$2$ ancestors in one deme and at most one in each of the others come next (subspace $\tilde{S}_2$), and all the others come last (subspace $\tilde{S}_3$).
The transition matrix of the genealogical process starting from a sample of size $n$ can be written as
 \begin{equation}\label{}
 \mathbf{Q}_{}=\mathbf{A}(N)+\frac{\mathbf{B}(N)}{ND}+ \frac{\mathbf{O}(1)}{ND^2}, 
 \end{equation}
 where $\mathbf{A}(N)$ is the transition matrix in the case of an infinite number of demes. With respect to the state subspaces $\tilde{S}_1, \tilde{S}_2, \tilde{S}_3$ in this order, we have
 \begin{equation}\label{}
  \mathbf{P}(N)=\lim_{\tau \rightarrow\infty}\mathbf{A}(N)^{\tau}= \begin{pmatrix}
\mathbf{I}&\mathbf{0}&\mathbf{0} \\
\mathbf{P}_{21}(N)&\mathbf{0}&\mathbf{0} \\
\mathbf{P}_{31}(N)&\mathbf{0}&\mathbf{0} \\
\end{pmatrix},
\end{equation}
where $\mathbf{I}$ is an identity matrix, while the entries  of $ \mathbf{P}_{21}(N)$ and $\mathbf{P}_{31}(N)$ are probabilities of fixation from states in $S_2$ and $S_3$, respectively, to states in $S_1$ in the case of an infinite number of demes. Note that the non-null entries of $ \mathbf{P}_{21}(N)$ are either 
 \begin{equation}\label{}
 \tilde{f}_{21}=\frac{(1-m)^2/N}{(1-m)^2/N + 1- (1-m)^2}=\frac{(1-m)^2}{(1-m)^2 + Nm(2-m)}
 \end{equation}
 or
\begin{equation}\label{}
 \tilde{f}_{22}=\frac{1-(1-m)^2}{(1-m)^2/N + 1-(1-m)^2}=\frac{Nm(2-m)}{(1-m)^2 + Nm(2-m)},
 \end{equation}
 which are the probabilities for two individuals in the same deme, or two offspring in the same deme after dispersal, to have $1$ or $2$ ancestors, respectively, in different demes. Moreover, with respect to the same subspaces, we have
 \begin{equation}\label{}
  \mathbf{B}(N)=\begin{pmatrix}
\mathbf{B}_{11}(N)&\mathbf{B}_{12}(N)&\mathbf{0} \\
\mathbf{B}_{21}(N)&\mathbf{B}_{22}(N)&\mathbf{B}_{23}(N) \\
\mathbf{B}_{31}(N)&\mathbf{B}_{32}(N)&\mathbf{B}_{33}(N) \\
\end{pmatrix}.
\end{equation}
M\"ohle's (1998a \cite{*M1998a}; see Appendix B) lemma yields in this case
 \begin{equation}
\lim_{D\rightarrow\infty}\left( \mathbf{A}(N)+\frac{\mathbf{B}(N)}{ND}+\frac{\mathbf{O}(1)}{ND^2}\right)^{\lfloor tND\rfloor}=\mathbf{P}(N)\exp\{t\mathbf{G}(N)\},
\end{equation}
where 
 \begin{equation}\label{}
  \mathbf{G}(N)=\begin{pmatrix}
\mathbf{B}_{11}(N)+\mathbf{B}_{12}(N)\mathbf{P}_{21}(N)&\mathbf{0}&\mathbf{0} \\
\mathbf{P}_{21}(N)\mathbf{B}_{11}(N)+\mathbf{P}_{21}(N)\mathbf{B}_{12}\mathbf{P}_{21}(N)&\mathbf{0}&\mathbf{0} \\
\mathbf{P}_{31}(N)\mathbf{B}_{11}(N)+\mathbf{P}_{31}(N)\mathbf{B}_{12}\mathbf{P}_{21}(N)&\mathbf{0}&\mathbf{0} \\
\end{pmatrix}.
\end{equation}
This means that, after initial instantaneous transitions from states in $S_2$ and $S_3$ to states in $S_1$, the genealogical process with $ND$ time steps as unit of time as $D\rightarrow \infty$ is described by a continuous-time Markov chain on the state space $S_1$ with $\mathbf{B}_{11}(N)+\mathbf{B}_{12}(N)\mathbf{P}_{21}(N)=(b_{k,l})_{k,l=1}^n$ as infinitesimal generator with non-null entries given by
 \begin{equation}\label{}
b_{k,k-1}=-b_{k,k}=\frac{k(k-1)}{2}m(2-m) + \frac{k(k-1)}{2}m(2-m)(N-1)\tilde{f}_{21}=\frac{k(k-1)}{2}\tilde{f}_{22}
\end{equation}
for $k=2, \ldots, n$. Here, the elements of $S_1$ are represented by the number of ancestors in different demes. 

In conclusion, in the limit of a large number of demes $D$ with 
$ND/\tilde{f}_{22}$ time steps as unit of time, ancestral lines within demes either coalesce or migrate instantaneously to different demes and, once ancestral lines are in different demes, their number decreases by $1$ at the rate given by the number of pairs of lines, that is, $(k-1)k/2$ for $k=2, \ldots, n$. 

Note that the probability $\tilde{f}_{31}$ for three individuals in the same deme, or three offspring in the same deme after dispersal, to have only $1$ ancestor in the case of an infinite number of demes satisfies
\begin{equation}\label{}
 \tilde{f}_{31}=(1-m)^3\left(\frac{1}{N^2} + \frac{3}{N}\left( 1- \frac{1}{N}\right)\tilde{f}_{21} + \left(1-\frac{1}{N}\right)\left( 1- \frac{2}{N}\right)\tilde{f}_{31}\right),
 \end{equation}
 from which
 \begin{equation}\label{}
 \tilde{f}_{31}=\tilde{f}_{21}\left(\frac{Nm(1-m) + 2(N-1)(1-m)^3}{N^2m(3-3m+m^2)+(3N-2)(1-m)^3}\right).
 \end{equation}
Moreover, we have $\tilde{f}_{32}=3(\tilde{f}_{21}-\tilde{f}_{31})$, since there are three possibilities for two individuals out of three to have a common ancestor, and $\tilde{f}_{33}=1-\tilde{f}_{31}-\tilde{f}_{32}$. Here, $\tilde{f}_{32}$ and $\tilde{f}_{33}$ are 
 the probabilities for three individuals in the same deme, or three offspring in the same deme after dispersal, to have only $2$ and $3$ ancestors, respectively, in the case of an infinite number of demes. 

Note also that 
\begin{equation}\label{}
 \tilde{f}_{21}=(1-m)^2f_{21} + m(2-m)
  \end{equation}
  and
  \begin{equation}\label{}
 \tilde{f}_{31}=(1-m)^3f_{31} + 3m(1-m)^2f_{21} + m^2(3-2m),
  \end{equation}
where $f_{21} $ and $f_{31}$ are the corresponding probabilities for offspring in the same deme before dispersal. When $m$ is small and $N$ is large, we have the approximations
\begin{equation}\label{}
f_{21} \approx \tilde{f}_{21}\approx \frac{1}{1+2\nu}
  \end{equation}
  and
  \begin{equation}\label{}
f_{31} \approx \tilde{f}_{31}\approx \frac{1}{(1+\nu)(1+2\nu)}.
  \end{equation}
  Moreover, we have $f_{22}=1-f_{21}$, $f_{32}=3(f_{21}-f_{31})$ and $f_{33}=1-f_{31}-f_{32}$ for the other corresponding probabilities for offspring in the same deme before dispersal.

\section*{Acknowledgments} The author thanks D. Kroumi and J. Wakeley for their comments on an early draft of this paper. This research was supported in part by the Natural Sciences and Engineering Research Council of Canada (Discovery Grant no. 8833).

\end{document}